\documentclass[
	aps, prd, reprint,
	10pt, notitlepage, a4paper,
        floats, floatfix,
	amsmath, amssymb, amsfonts, eqsecnum,
	superscriptaddress,
	showpacs, showkeys,
	nofootinbib,
 	longbibliography,
]{revtex4-1}

\usepackage{graphicx} 
\usepackage[caption=false]{subfig}
\usepackage[usenames,dvipsnames]{xcolor}
\usepackage{xspace} 
\usepackage{bm} 
\usepackage[utf8]{inputenc} 

\xdefinecolor{mylinkcolor}{rgb}{0,0,0.5}
\usepackage[
	bookmarksnumbered, bookmarksopen, bookmarksopenlevel=2,
	breaklinks=true,
	colorlinks=true, filecolor=mylinkcolor, citecolor=mylinkcolor,
	linkcolor=mylinkcolor, urlcolor=mylinkcolor, menucolor=mylinkcolor,
]{hyperref}

\def\vct#1{{\bm{#1}}}

\newcommand{\nnm}{\nonumber}
\newcommand{\doe}{\partial}
\newcommand{\be}{\begin{equation}}
\newcommand{\ee}{\end{equation}}
\newcommand{\bea}{\begin{eqnarray}}
\newcommand{\eea}{\end{eqnarray}}
\newcommand{\bdm}{\begin{displaymath}}
\newcommand{\edm}{\end{displaymath}}
\newcommand{\bse}{\begin{subequations}}
\newcommand{\ese}{\end{subequations}}

\newcommand{\mr}{\mathrm}
\newcommand{\tr}{\textrm}
\newcommand{\mc}{\mathcal}

\newcommand{\bs}{\boldsymbol}

\DeclareMathOperator{\Order}{\mathcal{O}}

\allowdisplaybreaks[0]

\begin{document}

\newcommand{\AEI}{\affiliation{Max-Planck-Institute for Gravitational Physics (Albert-Einstein-Institute),
\\ Am M{\"u}hlenberg 1, 14476 Potsdam-Golm, Germany, EU}}

\title{Spin-multipole effects in binary black holes and the test-body limit}
\date{\today}
\author{Justin Vines}
\email{justin.vines@aei.mpg.de}
\AEI
\author{Jan Steinhoff} 
\email{jan.steinhoff@aei.mpg.de}
\homepage{http://jan-steinhoff.de/physics/}
\AEI

\begin{abstract}
We discuss the effects of the black holes' spin-multipole structure in the orbital
dynamics of binary black holes according to general relativity, focusing on the leading-post-Newtonian-order couplings at each order in an expansion in the black holes' spins.  We first review previous widely confirmed results up through fourth order in spin, observe suggestive patterns therein, and discuss how the results can be extrapolated to all orders in spin with minimal information from the test-body limit.  We then justify this extrapolation by providing a complete derivation within the post-Newtonian framework of a canonical Hamiltonian for a binary black hole, for generic orbits and spin orientations, which encompasses the leading post-Newtonian orders at all orders in spin.  At the considered orders, the results reveal a precise equivalence between arbitrary-mass-ratio two-spinning-black-hole dynamics and the motion of a test black hole in a Kerr spacetime, as well as an intriguing relationship to geodesic motion in a Kerr spacetime.
\end{abstract}

\maketitle

\section{Introduction}
Binary black holes (BBHs) [or something very much like them] have provided us with the first gravitational waves detected at Earth \cite{detectionpaper,BoxingDay,Abbott:2017vtc}.  So far, the detected signals lie within the error bars of our expectations---that black holes exist, that binaries of them emit gravitational waves which become ever stronger as they spiral into one another and eventually merge into one bigger black hole, and that all of this is governed by Einstein's theory of general relativity (GR) \cite{testingGR}.

\begin{figure}
\begin{center}
\includegraphics[scale=.6]{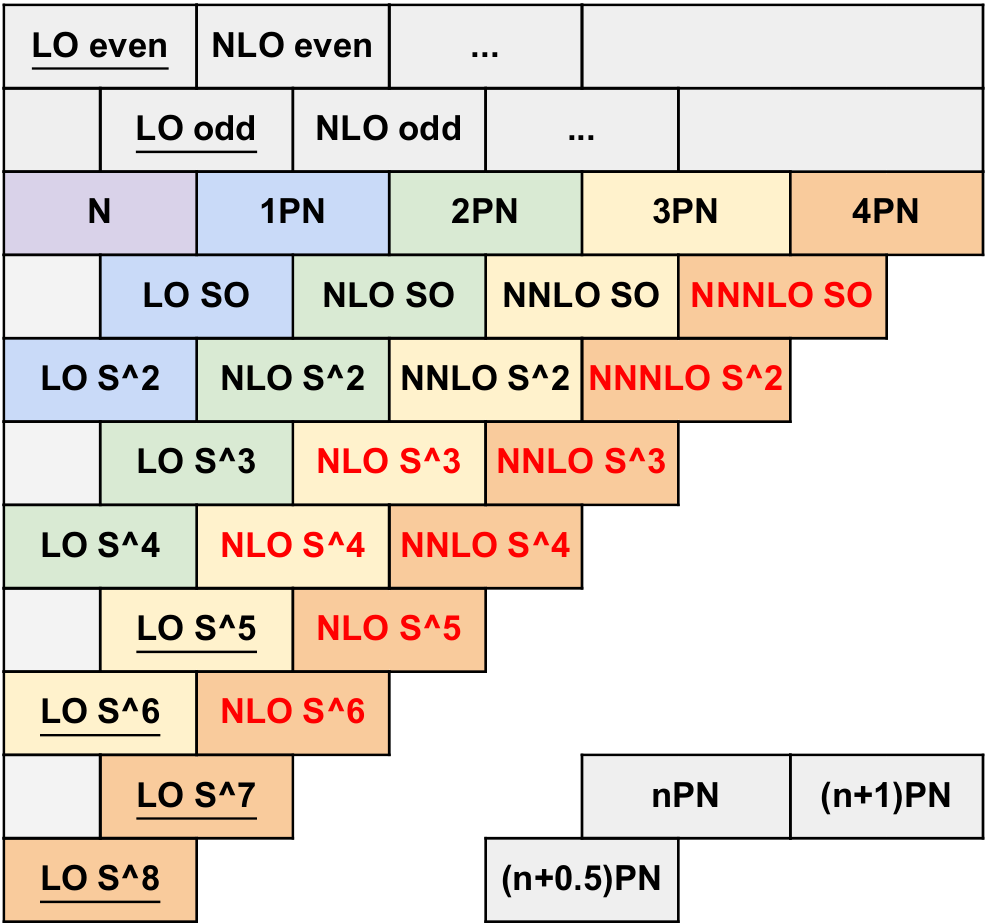}
\caption{Contributions to the two-body Hamiltonian in the PN-spin expansion, for arbitrary-mass-ratio binaries with spin-induced multipole moments (such as BBHs).  The lower-right inset indicates the increments in the PN orders in moving right or diagonally down and left; c.f.\ (\ref{schematicH}).
The first row has the nonspinning (point-mass) $S^0$ contributions, labelled by the order in the PN expansion: N for Newtonian, 1PN for relative order $\epsilon^1$, etc.  For the spin contributions, the second row gives the linear-in-spin or spin-orbit (SO) parts, the third row gives the quadratic-in-spin parts, etc.  LO stands for the leading-(PN-)order part (the $S^n$ contributions at the leading order in $\epsilon$), NLO stands for next-to-leading order, etc.
Terms in red text are unknown.  Terms in black text have been calculated, and confirmed by independent groups, as discussed at LOs below and as reviewed e.g.~in \cite{Blanchet:2006} at higher orders, all except for (i) the recent NNLO S$^2$ calculations of \cite{Levi:Steinhoff:2015:3}, and (ii) the underlined LO-S$^n$ terms with $n\ge 5$, which are presented here for BBHs. The recent 4PN results of \cite{Damour:2014jta, Damour:2015isa, Bernard:2015njp, Damour:2016abl, Bernard:2016wrg} have been recently also confirmed by independent calculations.}\label{fig}
\end{center}
\end{figure}

Being able to make such statements requires that we know exactly what it is that GR predicts.  Decades of work in numerical relativity (solving Einstein's equations directly on a supercomputer), combined with analytic approximation schemes for treating the two-body problem, have led to the current understanding of BBHs which has allowed the analysis of the detected signals.  Yet, in many respects, the relativistic two-body problem remains unsolved (see e.g.~Fig.~\ref{fig}).

For analytic attacks, two complementary approximation schemes are available: The post-Newtonian (PN) approximation expands about the Newtonian (weak-field, slow-motion) limit but is valid for arbitrary mass ratios~\cite{Blanchet:2006,poisson2014gravity}, while the extreme-mass-ratio approximation, encompassing the ``self-force paradigm'' \cite{Detweiler:2003ci,Poisson:Pound:Vega:2011,Barack:2014,Harte:review,Pound_review,Bini:2016cje,Bini:2016dvs}, expands about the test-body limit but is valid in the strong-field, relativistic regime.

This paper points out an interplay between the PN limit and the test-body limit, which relies on special properties (seemingly) specific to BBHs in GR, and which reveals structure in the BBH dynamics which has yet to be recognized.  We derive here a Hamiltonian for the conservative dynamics of generic-orbit arbitrary-mass-ratio spinning BBHs, at the leading orders in the large-separation/slow-motion (PN) expansion (at each order in spins), to all orders in the black holes' spins.  We find that these leading-order couplings can all be obtained from a map to the motion of a test black hole (a test body with the spin-induced multipoles of a Kerr black hole) in a background Kerr spacetime [see Eqs.~(\ref{testmap}) and (\ref{spinmap}) below], as is confirmed with direct post-Newtonian calculations for arbitrary mass ratios.  Furthermore, all of the couplings can be ``deduced'' in a certain manner from those of a pole-dipole test body in a Kerr spacetime.

The leading-PN-order, all-orders-in-spin BBH Hamiltonian is given explicitly by (\ref{Hfinal}) below.  This result (with its conceivable generalizations beyond leading order) is of particular interest for the case of large black hole spins, and is thus relevant to LIGO's ability to test GR with strong, nonlinear spin/precession effects in BBHs \cite{testingGR}.  

It is remarkable that the nonperturbative structure emerging in (\ref{Hfinal}) is an oblate spheroidal geometry (as in Fig.~\ref{fig:OS}) with a ring-radius being the sum of the individual black holes' ring-radii; this points to a spinning analog of the fact that Newtonian two-body dynamics is equivalent to the motion of a reduced mass in the gravitational potential of the sum of the individual masses.

The paper is organized as follows. In Sec.~\ref{BBHPN} we review PN results for spinning BBH dynamics
and extrapolate to the main result of the paper.   A proof of this result is given in Sec.~\ref{derivation}, where we derive the leading-order PN interaction potential for a BBH
at each order in the spins and make the connection to the
test-body limit. Our conclusions are given in Sec.~\ref{conclude}.

\section{BBH in the PN approximation}\label{BBHPN}
We can begin to explain and substantiate the above claims by reviewing the results of PN calculations which describe the conservative dynamics of binaries of compact objects with spin-induced multipole moments.

In the PN approximation, one describes a binary of compact objects, bodies $A=1,2$, in terms of
\begin{itemize}
\item their worldlines $\bs x=\bs z_A(t)$ in a PN spacetime \\with coordinates $x^\mu=(t,x^i)=(t,\bs x)$,
\\defining the relative position $\bs R=\bs z_1-\bs z_2$ \\and distance $R=|\bs R|$,
\item their masses $m_A$,
defining the total mass $M=m_1+m_2$, the reduced mass $\mu=m_1m_2/M$, and the symmetric mass ratio $\nu=\mu/M$,
taking $m_1\ge m_2$, with the ``test-body limit'' defined by $m_2\to 0$,
\item their intrinsic angular momentum (or spin) vectors $\bs S_A=S_A^i$,
defining the rescaled spin vectors \\$\bs a_A=\bs S_A/m_Ac$ with dimensions of length \\
(sometimes also referred to as the spins),

\item and their higher-order multipole moments,
\\beginning with the mass quadrupole tensors $Q_A^{ij}$,
\end{itemize}
with appropriate definitions for these quantities, sufficient for our purposes here, given e.g.\ in \cite{DSX1,DSX2} or \cite{RF}.

In general, the bodies' quadrupoles and higher-order moments can be dynamical, depending on further internal degrees of freedom \cite{Flanagan:2007ix,Chakrabarti:2013xza,Goldberger:2005cd,Hinderer:2016eia}.  But the leading-order effects in the post-Newtonian regime arise from \mbox{(i) intrinsic spin-induced} quadrupolar deformations scaling as the square of the spin, and \mbox{(ii) quadrupolar tidal deformations} which are adiabatically induced by the external field, which contribute to the quadrupole tensors as follows \cite{Poisson:1997ha},
\be\label{introQ}
Q_A^{ij}=-\kappa_Am_A a_A^{<i}a_A^{j>}-\lambda_A \mc E^{ij}(\bs z_A).
\ee
Here, $\mc E_{ij}$ is the electric tidal tensor, with $\mc E_{ij}=\doe_i\doe_j\phi$ in the Newtonian limit, $\phi$ being the Newtonian potential with $\phi=-Gm/R$ for a monopole.  Anglular brackets denote symmetric-trace-free (STF) projection.

The linear response coefficients $\kappa_{A}$ and $\lambda_{A}$ measure the leading-order quadrupolar deformation of body $A$, due to its rotation/spin and due to the external tidal field, respectively.  The values appropriate for a black hole,
\be
\kappa_\mr{BH}=1,\qquad \lambda_\mr{BH}=0,
\ee
have been established through several arguments and derivations, e.g.~\cite{Hansen:1974zz,Poisson:1997ha,Porto:Rothstein:2008:2, Taylor:2008xy,Damour:2009vw,Kol:2011vg,Pani:2015hfa}.

We will henceforth restrict attention to spin-induced multipoles, as is appropriate for black holes at the PN orders discussed here.  
Spinning bodies (like black holes) generally have
\begin{itemize} 
\item even-order mass multipoles $\mc I^L$,

quadrupole $Q^{ij}\equiv\mc I^{ij}$,  hexadecapole $\mc I^{ijkl}$, 
\\\dots, $2^\ell$-pole $\mc I^L=\mc I^{i_1\ldots i_{\ell}}$ with $\ell$ even, and

\item odd-order current multipoles $\mc J^L$,

dipole $S^i\equiv\mc J^i$, octupole $\mc J^{ijk}$, \\\ldots, $2^\ell$-pole  $\mc J^L=\mc J^{i_1\ldots i_{\ell}}$ with $\ell$ odd,
\end{itemize}
which are induced by their rotation, and which are generally proportional to STF outer products of $\ell$ copies of the spin vector.  $L=i_1\ldots i_\ell$ is a spatial multi-index.  The proportionality constants (like $\kappa$ for $\ell=2$) vary with the composition and structure of the bodies.  The multipoles of a black hole (with certain normalizations),  including the mass monopole $m\equiv\mc I$, are given by \cite{Hansen:1974zz}
\be\label{IJBH}
\bigg(\mc I^L+\frac{i}{c}\mc J^L\bigg)_\mr{BH}\,=\,i^\ell\,m\, a^{<L>},
\ee
where $a^{<L>}=a^{<i_1}\ldots a^{ i_\ell>}=S^{<L>}/(mc)^\ell$, with $\ell=0,1,\ldots,\infty$. With $\ell=2$, this reproduces (\ref{introQ}) with $\kappa=1$ (and $\lambda=0$).

With only such spin-induced multipoles, the only degrees of freedom of the binary are the relative position $\bs R(t)$ (in the center-of-mass frame) and the spins $\bs S_1(t)$ and $\bs S_2(t)$.  The PN
conservative dynamics can be encoded in a Hamiltonian $H(\bs R, \bs P, \bs S_1,\bs S_2)$, where $\bs P$ is the linear momentum canonically conjugate to $\bs R$, and the equations of motion are determined from
\be\label{Heoms}
\dot R^i=\frac{\doe H}{\doe P_i},\quad \dot P_i=-\frac{\doe H}{\doe R^i},\quad \dot S_A^i=\epsilon^{ij}{}_k\frac{\doe H}{\doe S_A^j}S_A^k,
\ee
with $A=1,2$.

The Hamiltonian, expanded in PN orders and in powers of the spins, takes the following form, $H=$
\begin{align}\label{schematicH}
&\phantom{+}H^\mr{(0PN)}_{\tr{LO-S}^0}
\\
&\!+H^\mr{(1PN)}_\mr{\tr{NLO-S}^0}+H^\mr{(1.5PN)}_{\tr{LO-S}^1}+H^\mr{(2PN)}_{\tr{LO-S}^2}\phantom{\bigg|}
\nnm\\
&\!+H^\mr{(2PN)}_\mr{\tr{NNLO-S}^0}+H^\mr{(2.5PN)}_{\tr{NLO-S}^1}+H^\mr{(3PN)}_{\tr{NLO-S}^2}+H^\mr{(3.5PN)}_{\tr{LO-S}^3}+H^\mr{(4PN)}_{\tr{LO-S}^4}
\nnm
\end{align}
$+\ldots$ Here, $H_{\tr{LO-S}^0}\equiv H_\mr{N}$ is the Newtonian (0PN) point-mass (no-spin) Hamiltonian, and the other terms are at higher orders in the PN parameter $\epsilon\sim Gm/c^2R\sim v^2/c^2$ and higher orders in the spin.  The order counting here, with an $n$PN term scaling as $\epsilon^nH_\mr{N}$, assumes rapidly rotating bodies, with spin magnitudes  $S\sim Gm^2/c$.

We now discuss in turn the contributions in (\ref{schematicH}).  To ease the notation, we henceforth set $G=c=1$ and define new momenta rescaled by the reduced mass,
\be\label{rescale}
\bar H=\frac{H}{\mu},\qquad \bar{\bs P}=\frac{\bs P}{\mu},\qquad \bar{\bs L}=\frac{\bs L}{\mu}=\bs R\times\bar{\bs P},
\ee
where $\bs L=\bs R\times\bs P$ is the orbital angular momentum.

\subsection{Nonspinning}

The Newtonian Hamiltonian [i.e.\ the leading-order (LO) no-spin (S$^0$) Hamiltonian] reads
\be
\bar H_\mr{N}=\frac{\bar{\bs P}^2}{2}-\frac{M}{R},
\ee
and the 1PN point-mass Hamiltonian [i.e.\ the next-to-leading-order (NLO) no-spin Hamiltonian], in harmonic/ADM gauge, reads
\begin{align}
H_\mr{\tr{NLO-S}^0}&=(-1+3\nu)\frac{\bar{\bs P}^4}{8}+(-3-2\nu)\frac{M\bar{\bs P}^2}{2R}
\nnm\\
&\quad+(0+\nu)\frac{M\bar{\bs L}^2}{2R^3}+(1+0\nu)\frac{M^2}{2R^2}.
\end{align}

The Newtonian Hamiltonian could be said to be equal to its test-body limit, in the sense that it has no dependence on the mass ratio.  The same is not true of $H_\mr{1PN}$, and one recovers only the first terms in parentheses from the test-body limit, with the terms $\propto\nu$ being ``self-force corrections'' \cite{Detweiler:2003ci,Poisson:Pound:Vega:2011,Barack:2014,Harte:review,Pound_review}.

More precisely, even $H_\mr{N}$ is not literally equal to its test-body limit, if this is defined as the limit $m_2\to0$, because then we obtain only the $m_1$ term in $M=m_1+m_2$ [among other subtleties].  Rather, the arbitrary-mass-ratio Newtonian dynamics in the center of mass frame is \emph{equivalent} to the dynamics of a test body with mass $\mu$ in the field of a stationary mass $M$, under the map $M=m_1+m_2$ and $\mu=m_1m_2/M$.  This map and this fortuitous coincidence allow us to obtain from the test-body limit what could be considered a self-force correction (the $m_2$ in $M$) [and less handwavingly, the full exact Newtonian dynamics].  We will find this same kind of coincidence to all orders in spin at the leading PN orders.

\subsection{Leading-order spin-orbit couplings}

Next we have the leading-order ``spin-orbit'' (linear-in-spin) Hamiltonian, at 1.5PN, \cite{Tulczyjew:1959,D'Eath:1975vw,PhysRevD.2.1428,Damour:1988mr}
\begin{align}
\bar H_{\tr{LO-S}^1}&=\left[2m_1+\frac{3}{2}m_2\right]\frac{\bar{\bs L}\cdot\bs a_1}{R^3}+\left[\frac{3}{2}m_1+2m_2\right]\frac{\bar{\bs L}\cdot\bs a_2}{R^3}.
\end{align}
The $m_2$ terms are self-force corrections which drop out in the test-body limit.  But under the map
\begin{align}\label{spinmap}
\bs S=\bs S_1+\bs S_2&=m_1\bs a_1+m_2\bs a_2=M\bs \sigma,
\nnm\\
\frac{\bs S_\mr{test}}{\nu}=\bs S^*=\frac{m_1}{m_2}\bs S_2+\frac{m_2}{m_1}\bs S_1&=m_1\bs a_2+m_2\bs a_1=M\bs\sigma^*,\phantom{\bigg|}
\end{align}
this can be rewritten exactly as the LO linear-in-spins part of the Hamiltonian of a test body with mass $\mu$ and spin $\bs S_\mr{test}=\mu\bs\sigma^*=\nu\bs S^*$ in the field of a stationary body with mass $M$ and spin $\bs S=M\bs\sigma$,
\begin{align}\label{SOfinal}
\bar H_{\tr{LO-S}^1}(m_1,\bs a_1,m_2,\bs a_2)
&=\bar H^\mr{test}_{\tr{LO-S}^1}(M,\bs \sigma,\mu,\bs\sigma^*)
\\\nnm
&=\bar{\bs L}\cdot\left(2\bs\sigma+\frac{3}{2}\bs \sigma^*\right)\frac{M}{R^3}
\\\nnm
&=-\bar{\bs P}\times\left(2\bs\sigma+\frac{3}{2}\bs \sigma^*\right)\cdot\bs\nabla\, \frac{M}{R},
\end{align}
where $\bs\nabla=\doe_i=\doe/\doe R^i$.

Note that the dynamics defined by $H_\mr{N}$, $H_\mr{1PN}$, and $H_{\tr{LO-S}^1}$ (and more generally, through linear order in spin) is universal, independent of the nature of the bodies.  

\subsection{Leading-order spin-squared couplings}

The next contribution in (\ref{schematicH}) is the LO-S$^2$ Hamiltonian at 2PN, \cite{D'Eath:1975vw,Barker:1975ae,Barker1979,Damour1SEOB,Racine:2008qv}
\begin{align}
\bar H_{\tr{LO-S}^2}&=\frac{1}{2}\Big(\kappa_1a_1^ia_1^j+2a_1^ia_2^j+\kappa_2a_2^ia_2^j\Big)\doe_{i}\doe_{j}\frac{M}{R},
\end{align}
which begins to depend on the bodies' internal structure through the response coefficients $\kappa_{1,2}$.  Note that the $\kappa$ terms encode the coupling of the spin-induced quadrupole of one body to the monopole of the other, while the $a_1$-$a_2$ term encodes the (universal) coupling between the bodies' spins.

Remarkably, for the special case of a binary black hole, $\kappa_1=\kappa_2=1$, this factorizes into \cite{Damour1SEOB}
\begin{alignat}{3}
&\bar H_{\tr{LO-S}^2}^\tr{BBH}(m_1,\bs a_1,m_2,\bs a_2)=\frac{1}{2}\big(\bs a_1+\bs a_2\big)^i\big(\bs a_1+\bs a_2\big)^j\doe_{i}\doe_{j}\frac{M}{R}
\nnm\\\label{onehandS2}
&=\bar H_{\tr{LO-S}^2}^\tr{BBH,test}(M,\bs\sigma,\mu,\bs\sigma^*)=\frac{1}{2}\big((\bs\sigma+\bs\sigma^*)\cdot\bs\nabla\big)^2\frac{M}{R}
\\\label{otherhandS2}
&=\bar H_{\tr{LO-S}^2}^\tr{BBH,test}(M,\bs a_0,\mu,0)=\frac{1}{2}(\bs a_0\cdot\bs\nabla)^2\frac{M}{R},
\end{alignat}
noting that spin vectors commute with spatial derivatives $\bs\nabla$, and where
\begin{align}\label{a0}
\bs a_0&=\bs a_1+\bs a_2=\bs\sigma+\bs \sigma^*
=\frac{\bs S+\bs S^*}{M}=\frac{\bs S_0}{M}
\end{align}
is the combination of the spins whose importance was noted in \cite{Damour1SEOB,DJS,Racine:2008qv}.

The LO-S$^2$ BBH Hamiltonian is equivalent to that of a test-body in two different ways.  On the one hand, as in (\ref{onehandS2}), it is the LO quadratic-in-spins part of the Hamiltonian of a ``test black hole'' with mass $\mu$ and spin $\mu\bs\sigma^*$ (and quadrupole $Q_{ij}=-\mu\sigma^*_{<i}\sigma^*_{j>}$) in the field of a stationary Kerr black hole with mass $M$ and spin $M\bs\sigma$.  On the other hand, as in (\ref{otherhandS2}), it is the LO-S$^2$ part of the Hamiltonian of a structureless point mass (following a geodesic) in the field of a Kerr black hole with mass $M$ and spin $M\bs a_0$ \cite{Damour1SEOB,DJS}.

\subsection{Leading-order couplings for binary black holes through fourth order in spin}

The LO-S$^3$ (3.5PN) and LO-S$^4$ (4PN) contributions in (\ref{schematicH}) have been computed and confirmed by a variety of methods in \cite{  Hergt:2007ha, Hergt:2008jn, Levi:Steinhoff:2014:2,Vaidya:2014kza, Marsat:2014xea}.  To the authors' knowledge, there are no previous results for the PN dynamics of arbitrary-mass-ratio binaries at fifth or sixth order in the spins (5.5PN or 6PN at LO) or beyond, though much is known (at least in principle) from the test body limit.

The LO-S$^3$ contributions arise from (i) a body's spin-induced current octupole coupling to its companion's mass monopole, (ii) the mass quadrupole coupling to the companion's spin, and (iii) more subtle kinematical effects.  These kinematical effects, like those encountered for the spin-orbit couplings (\ref{SOfinal}) which are linked to Thomas precession \cite{Damour:1991rd}, are related to the transport of the local frame in which the spin is defined and its interplay with the spin supplementary condition \cite{Levi:Steinhoff:2015:1}.

The LO-S$^4$ contributions arise from hexadecapole-monopole, octupole-dipole, and quadrupole-quadrupole couplings.  As with the LO-S$^2$ couplings, there is no dependence on $\bar{\bs P}$, only on $\bs R$, and there are no subtle kinematical effects.

Like the LO-S$^2$ part, the LO-S$^3$ and LO-S$^4$ parts undergo remarkable simplifications in the special case when the spin-induced multipole moments match those of a Kerr black hole.

Now we gather all the results for the leading-PN-order Hamiltonians at each order in spin, available from \cite{  Hergt:2007ha, Hergt:2008jn, Levi:Steinhoff:2014:2, Marsat:2014xea,Vaidya:2014kza} through fourth order in spin, specializing to the BBH case.  This is as in (\ref{schematicH}), but where we will neglect the NLO terms $H_\mr{1PN}$, $H_{\tr{NLO-S}^1}$, $H_{\tr{NLO-S}^2}$ and (at NNLO) $H_\mr{2PN}$, as well as all other NLO terms.  Working from the Hamiltonians of \cite{Levi:Steinhoff:2014:2}, after a canonical transformation affecting only the $S^3$ terms, and after some simplification, using (\ref{spinmap}), the leading-order Hamiltonian can be written as $\bar H^\mr{BBH}_\tr{LO}=\bar H^\mr{BBH}_\tr{LO,even}+\bar H^\mr{BBH}_\tr{LO,odd}$, with the even-in-spins part
\begin{align}\label{BBHeven4}
\bar H^\mr{BBH}_\tr{LO,even}=\frac{\bar{\bs P}^2}{2}-\frac{M}{R}
&+\frac{1}{2!}(\bs a_0\cdot\bs\nabla)^2\,\frac{M}{R}
\\\nnm&
-\frac{1}{4!}(\bs a_0\cdot\bs\nabla)^4\,\frac{M}{R}
+\mc O(S^6),
\end{align}
and the odd-in-spins part
\begin{alignat}{3}\label{BBHeven3}
\bar H^\mr{BBH}_\tr{LO,odd}&=-\frac{1}{1!}\bar{\bs P}\times\!\left[2\bs\sigma+\frac{3}{2}\bs \sigma^*\!\right]\!\cdot\bs\nabla\, \frac{M}{R}
\\\nnm
&\quad+\frac{1}{3!}\bar{\bs P}\times\!\left[2\bs\sigma+\frac{1}{2}\bs \sigma^*\!\right]\!\cdot\!\bs\nabla(\bs a_0\!\cdot\!\bs\nabla)^2 \frac{M}{R}+\mc O(S^5).
\end{alignat}
Note that hidden within these ``factorized'' forms is a considerable network of multipole-multipole couplings and kinematical effects, including ``(first-order) self-force \mbox{corrections}.''

All of these LO-PN results (for arbitrary mass ratios), even and odd, are obtained from the test-body limit according to
\be\label{testmap}
\bar H^\mr{BBH}_\tr{LO}(m_1,\bs a_1,m_2,\bs a_2)=\bar H^\mr{BBH,test}_\tr{LO}(M,\bs\sigma,\mu,\bs\sigma^*),
\ee
with (\ref{spinmap}), where $\bar H^\mr{BBH,test}$ is the Hamiltonian of a ``\emph{test black hole}'' with mass $\mu$ and spin $\mu\bs\sigma^*$---having all of the spin-induced multipoles of a black hole, keeping $\bs\sigma^*$ finite as $\mu\to 0$, noting that all of the LO couplings end up with one factor of $\mu$ which scales away as in (\ref{rescale})---in a Kerr spacetime with mass $M$ and spin $M\bs\sigma$.
The even part has the further feature
\be\label{a0map}
\bar H^\mr{BBH}_\tr{LO,even}(m_1,\bs a_1,m_2,\bs a_2)=\bar H^\mr{BBH,test}_\tr{LO,even}(M,\bs a_0,\mu,0),
\ee
so that it is obtained from geodesic motion in a Kerr spacetime with mass $M$ and spin $M\bs a_0$; this is not true of the odd part.  These hold for the above results through fourth order in spin, as can be confirmed from \cite{VKSH}, and we will see that they hold to all orders.  Note that this means that an effective-one-body Hamiltonian which uses $M\bs a_0$ as the spin for an effective ($\nu$-deformed) Kerr metric entering the geodesic Hamiltonian (a recent example being that in \cite{Balmelli:Damour:NLOSS}) correctly encodes all of the LO even-in-spin couplings.

\subsection{To all orders in spin, at the leading post-Newtonian orders, for binary black holes}

There is a clear pattern developing in the even part (\ref{BBHeven4}).  In light of (\ref{IJBH}), one is motivated to argue that this pattern continues,
\begin{align}\label{LOBBHeven}
\bar H^\mr{BBH}_\tr{LO,even}&=\frac{\bar{\bs P}^2}{2}-\sum_\ell^\mr{even}\frac{i^\ell}{\ell!}(\bs a_0\cdot\bs\nabla)^\ell\,\frac{M}{R}
\\\nnm
&=\frac{\bar{\bs P}^2}{2}-\cos(\bs a_0\cdot\bs\nabla)\,\frac{M}{R}.
\end{align}

For the odd part, one could argue from the limited data in (\ref{BBHeven3}) that there is an analogous pattern developing, with only two new coefficients at each order in spin (the coefficients of $\bs\sigma$ and $\bs\sigma^*$ in the cross product).  If we were to assume that this pattern holds to all orders, then these coefficients would all be fixed by matching to the dynamics of a pole-dipole test body in the Kerr spacetime, obeying the Mathisson-Papapetrou-Dixon (MPD) equations \cite{Mathisson:1937, Papapetrou:1951pa, Dixon:1979} to linear order in the spin of the test body.  The resultant coefficients are available in principle from \cite{Barausse:Racine:Buonanno:2009,VKSH}. 
This yields
\begin{align}\label{LOBBHodd}
&\bar H^\mr{BBH}_\tr{LO,odd}
\nnm\\\nnm
&=\sum_\ell^\mr{odd}\frac{i^{\ell-1}}{\ell!}\bar{\bs P}\times\left(-2\bs\sigma+\frac{\ell-4}{2}\bs \sigma^*\right)\cdot\bs\nabla\;(\bs a_0\cdot\bs\nabla)^{\ell-1}\, \frac{M}{R}
\\\nnm
&=\bigg[-2\,\bar{\bs P}\times\bs a_0\cdot\bs\nabla\,\frac{\sin(\bs a_0\cdot\bs\nabla)}{\bs a_0\cdot\bs\nabla}
\\&\qquad+\frac{1}{2}\,\bar{\bs P}\times\bs\sigma^*\cdot\bs\nabla\,\cos(\bs a_0\cdot\bs\nabla)\bigg]\frac{M}{R}.
\end{align}

These results can also be derived without relying on such seemingly unjustified extrapolation.  
With a direct PN calculation for arbitrary mass ratios in Sec.~\ref{derivation}, 
we show that the LO BBH Hamiltonian is indeed given by the sum of (\ref{LOBBHeven}) and (\ref{LOBBHodd}) under the map (\ref{spinmap}).

 The calculations of Sec.~\ref{derivation} are based on the well-developed action description of spinning bodies in general relativity \cite{Hanson:Regge:1974,Bailey1975,Bailey:Israel:1980,Porto:2005ac, Porto:2007qi,Porto:2008tb,Porto:Rothstein:2008:2,Levi:2008nh, Steinhoff:2008, Levi:2010zu,Porto:2010tr,Perrodin:2010dy,Levi:Steinhoff:2015:1,PORTO20161}, which results in a form of the MPD dynamics \cite{Mathisson:1937, Papapetrou:1951pa, Dixon:1979}.  The action encodes both the bodies' motion in an effective external field and the effective stress-energy which sources the field equations---at the least, in the leading-PN-order context.  In Sec.~\ref{derivation}, we draw in particular from the analyses of spin-multipole effects in \cite{Steinhoff:2011,Marsat:2014xea,Steinhoff:2015,Bohe:2015ana,VKSH}, and \cite{Levi:Steinhoff:2015:1} which derived all LO spin-induced multipole couplings.  We obtain all needed coupling constants in the action by matching the effective black holes' spin-multipole structure to that of a Kerr black hole, and by ensuring the kinematical consistency of the MPD dynamics.

\begin{figure}
\begin{center}
\includegraphics[scale=.45]{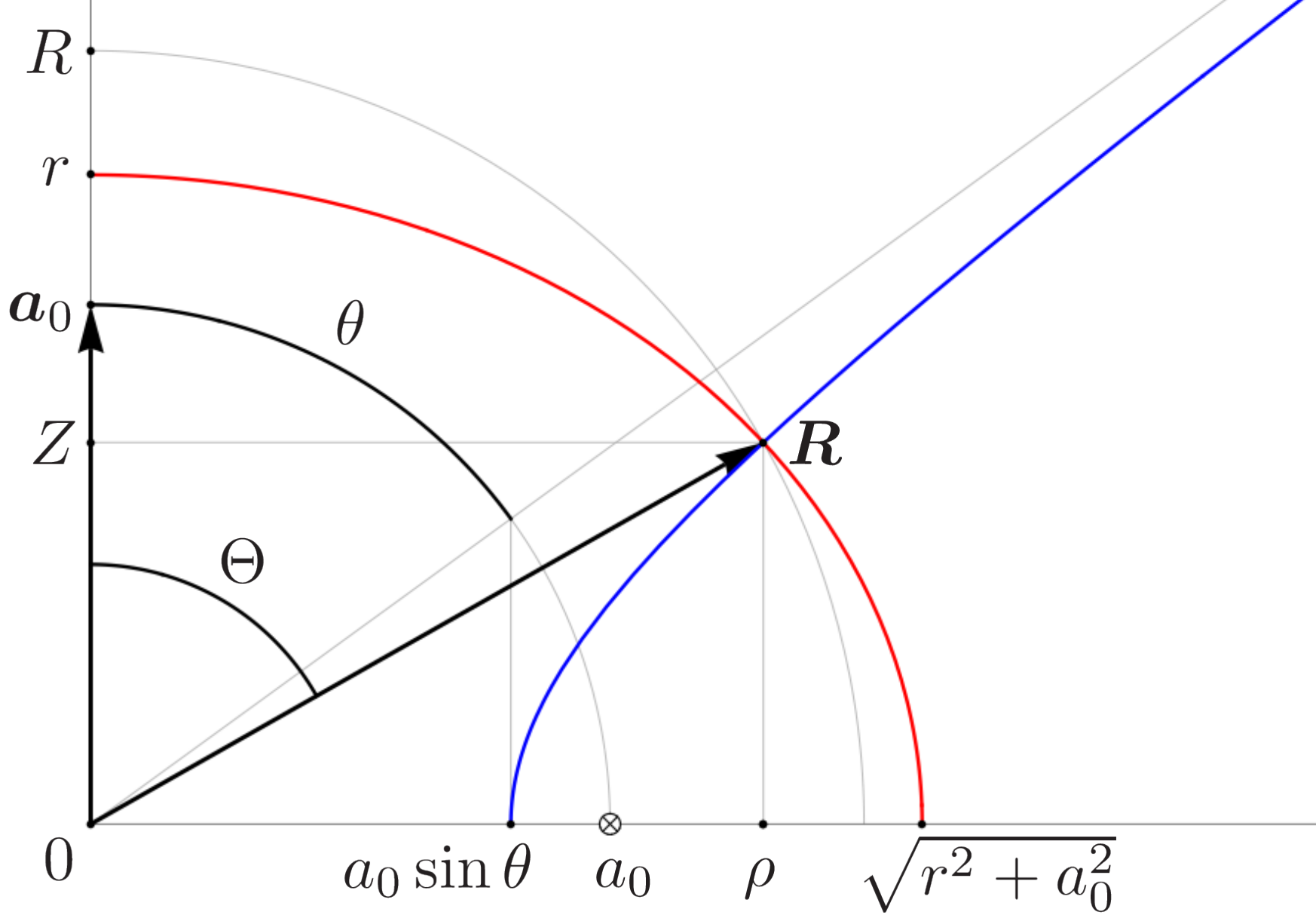}
\caption{Relationships between the coordinates (\ref{coords}) on flat 3-space, with $(r,\theta,\Phi)$ being oblate spheroidal coordinates with ``ring-radius'' $a_0=|\bs a_0|$, showing a quadrant of a plane containing the $Z$-axis. The surfaces of constant $r$ are oblate ellipsoids with foci on ``the ring'' $\rho^2=X^2+Y^2=a_0^2$ in the $Z=0$ plane, with a cross-section shown in red.  The ring pierces this plane orthogonally at the $\otimes$ symbol, with its center at the origin.  The locus $r=0$ is the disk $Z=0$, $\rho<a_0$ bounded by the ring.  The surfaces of constant $\theta$ are half- one-sheeted hyperboloids with foci on the same ring, with a cross-section shown in blue; as $r\to\infty$, they asymptote to cones opening an angle $\theta$ from the +$Z$-axis.  The locus $\theta=\pi/2$ is the plane $Z=0$ minus the disk $\rho<a_0$.  The ring represents the ``ring singularity'' of an effective Kerr black hole with rescaled spin $\bs a_0$.}\label{fig:OS}
\end{center}
\end{figure}

We can also present the results (\ref{LOBBHeven}) and (\ref{LOBBHodd}) in a simple explicit closed form by introducing new coordinates on the (flat) 3-space; starting from Cartesian coordinates $(X,Y,Z)$, as illustrated in Fig.~\ref{fig:OS}, we have
\begin{align}\label{coords}
&\tr{{cylindrical}}\; (\rho,\Phi,Z),\quad
X=\rho \cos\Phi,\phantom{\Big|}\;\; Y=\rho\sin\Phi,
\nnm\\\nnm
&\tr{{spherical}}\;(R,\Theta,\Phi),\quad
\rho=R\sin\Theta,\quad Z=R\cos\Theta,
\\
&\tr{{spheroidal}}\;(r,\theta,\Phi),\quad
\rho=\sqrt{r^2+a_0^2}\,\sin\theta,\quad Z=r\cos\theta.
\end{align}

The LO Hamiltonian, the sum of (\ref{LOBBHeven}) and (\ref{LOBBHodd}), can then be written as
\begin{align}\label{Hfinal}
H^\mr{BBH}_\mr{LO}
&\;=\;\frac{{\bs P}^2}{2\mu}\;+\;\mu \phi\;+\;  \bs P\cdot\bs A\;-\;\frac{1}{2}\bs P\times\bs\sigma^*\cdot\bs\nabla \phi\,,
\end{align}
where
\be\label{phi0}
\phi\,=\,-\cos(\bs a_0\cdot\bs\nabla)\frac{M}{R}\,=\,-\frac{Mr}{r^2+a_0^2\cos^2\theta}, 
\ee
and
\be\label{potentials}
\bs A\,=\,-2\bs a_0\times\bs\nabla\frac{\sin(\bs a_0\cdot\bs\nabla)}{\bs a_0\cdot\bs\nabla}\frac{M}{R}\,=\,2\phi\,\frac{\bs R\times\bs a_0}{r^2+a_0^2},
\ee
are the (linearized-harmonic-gauge) gravito-electric scalar potential $\phi=-h_{00}/2$ and gravito-magnetic vector potential $\bs A=-h_{0i}$ (with $h_{\mu\nu}=g_{\mu\nu}-\eta_{\mu\nu})$ of an effective Kerr black hole with mass $M$ and spin $M\bs a_0$,
and with
\be
\bs\nabla \phi=\dfrac{M}{2}\dfrac{\bs R+i\bs a_0}{(r+ia_0\cos\theta)^3}+\textrm{complex conjugate}.
\ee  
The manipulations linking (\ref{LOBBHeven})--(\ref{LOBBHodd}) to (\ref{Hfinal})--(\ref{potentials}) are similar to those in \cite{Israel:1970,Will:2008ys}.  

As we discuss in Sec.~\ref{sec:test}, the Hamiltonian given by (\ref{Hfinal})--(\ref{potentials}) is precisely the LO part of the Hamiltonian for a test black hole with mass $\mu$ and spin $\mu\bs\sigma^*$ in a Kerr spacetime with mass $M$ and spin $M\bs\sigma$.  We see that one obtains the arbitrary-mass-ratio BBH Hamiltonian from this simply by using the mass and spin mappings discussed above---i.e., the equivalence stated in (\ref{testmap}) holds to all orders in spin.

The relationship between the even-in-spin part and geodesics in a Kerr spacetime with spin $M\bs a_0$, stated in (\ref{a0map}), is seen to hold to all orders in spin as well---and (\ref{Hfinal})--(\ref{potentials}) also show some relationship between the odd-in-spin part and the Kerr spacetime with spin $M\bs a_0$.  We elaborate on these points and specify the precise relationship between the Kerr metric and the potentials (\ref{phi0})--(\ref{potentials}) also in Sec.~\ref{sec:test}.

It is astounding that the doubly infinite web of multipole-multipole couplings and kinematic effects present in the Hamiltonians---with all powers of both the test-black-hole spin $\bs\sigma^*$ and the Kerr spin $\bs\sigma$, or equivalently, with all powers of $\bs a_1$ and $\bs a_2$---reduces to the compact form (\ref{Hfinal}), depending in a simple way on the multipole structure of a single Kerr spacetime with spin $\bs a_0$.

\section{Derivation of the result}\label{derivation}
In this section we derive the result (\ref{Hfinal}) presented above, proceeding in the following steps.
We start by constructing the effective point-particle action 
for a rotating black hole (neglecting tidal effects). Next we approximate this action,
first to linear order in a perturbation of the metric away from flat spacetime, then
also to leading order in the velocity at each order in spin. Finally, we
compute the interaction potential of the binary to leading PN order at each order in spin.

\subsection{Effective action for black holes}

The effective point-particle action of a rotating black hole (neglecting tidal deformations)
can be written as an integral along an arbitrarily parametrized worldline $x^\mu=z^\mu(\sigma)$ as
\begin{equation}\label{SBH}
  \mathcal{S}_\text{BH} = \int d\sigma \left[ - m \sqrt{-u_\rho u^\rho} + L_\text{top} + L_\text{SI} \right] ,
\end{equation}
where the signature of spacetime is $+2$,
$u^\mu = d z^\mu / d \sigma$ is the worldline tangent, and $m$ is the mass, with the following Lagrangians:
$L_\text{top}$ describes a relativistic spinning spherical top minimally
coupled to gravity and $L_\text{SI}$ contains non-minimal couplings of the spin-induced (SI) multipole
moments.

We use the general-relativistic spherical-top Lagrangian in the form \cite{Levi:Steinhoff:2015:1}
\begin{equation}\label{Ltop}
L_\text{top} = S_{\mu\nu} \left[ \frac{1}{2} \Lambda_c{}^\mu \frac{D \Lambda^{c \nu}}{d \sigma}
    + U^\mu \frac{D U^\nu}{d \sigma} \right] ,
\end{equation}
where $U^\mu = u^\mu / \sqrt{-u_\rho u^\rho}$, $D$ is the covariant differential, and $\Lambda_c{}^\mu$
describes an ortho-normal (Lorentz) body-fixed frame, $g_{\mu\nu} \Lambda_a{}^\mu \Lambda_b{}^\nu = \eta_{ab}$.
The last term in Eq.~(\ref{Ltop}) is related to the Fermi-Walker transport
of the spin \cite{Misner:Thorne:Wheeler:1973}.
The spin 2-form $S_{\mu\nu} = - S_{\nu\mu}$ is subject to the constraint
\begin{equation}\label{SSC}
S_{\mu\nu} \left[ U^\nu + \Lambda_0{}^\nu \right] = 0 ,
\end{equation}
which must be supplemented by a gauge condition for $\Lambda_0{}^\mu$.
These conditions reduce the rotational degrees of freedom to the
three physical ones.\footnote{As discussed in \cite{Steinhoff:2015}, the action (\ref{SBH})--(\ref{Ltop}) uses dynamical variables associated to different instances of traditional spin supplementary conditions (SSCs) for the MPD equations \cite{Mathisson:1937, Papapetrou:1951pa, Dixon:1979}; such SSCs are discussed e.g.\ in \cite{Barker:1975ae,Barker1979,Kyrian:2007zz}.  The worldline $z^\mu$ here is that defined by the Tulczjew-Dixon SSC \cite{Tulczyjew:1959,Dixon:1979}, while the spin tensor $S^{\mu\nu}$ is that defined by the (thus-far) generic SSC defined by (\ref{SSC}), in which $\Lambda_0{}^\mu$ plays the role of a gauge field \cite{Steinhoff:2015}.} It is useful to define a spin vector as
\begin{equation}
S^\mu = m a^\mu = U_\nu *\!S^{\nu\mu} ,
\end{equation}
where the dual is defined as usual, $*S_{\mu\nu} = \frac{1}{2} \eta_{\mu\nu}{}^{\alpha\beta} S_{\alpha\beta}$ (with the volume form
$\eta_{\mu\nu\alpha\beta}$). The vector $a^\mu$
is normalized to the radius of the ring singularity of the rotating black hole
(the Kerr parameter). Both vectors are orthogonal to the motion,
$S^\mu U_\mu = 0 = a^\mu U_\mu$.

It is useful to introduce a tetrad field $e_a{}^\mu$ defining a
Lorentz frame at each spacetime point, $g_{\mu\nu} e_a{}^\mu e_b{}^\nu = \eta_{ab}$.
(Notice that $\Lambda_c{}^\mu$ is only defined on the worldline and encodes
the orientation of the body.) We use $e_a{}^\mu$ to translate between
spacetime indices $\mu,\nu,...$ and Lorentz indices $a,b,...$. The frame $e_a{}^\mu$
introduces six new gauge degrees of freedom (Lorentz transformations of the index $a$)
in addition to the four coordinate ones. Then the top Lagrangian reads
\begin{equation}
\begin{split}
L_\text{top} &=  S_{ab} \left[ \frac{1}{2} \Lambda_c{}^a \frac{d \Lambda^{c b}}{d \sigma}
  + U^a \frac{d U^b}{d \sigma} \right]
  + m *\!\omega_\mu{}^{ab} u^\mu U_a a_b  ,
\end{split}
\end{equation}
where we define the dual $*\omega_\mu{}^{ab} = \frac{1}{2} \eta^{ab}{}_{cd} \omega_\mu{}^{cd}$, the Ricci rotation coefficients $\omega_\mu{}^{ab} = e^b{}_\nu (\partial_\mu e^{a\nu} + \Gamma^\nu{}_{\alpha\mu} e^{a\alpha})$,
and the connection $2 \Gamma_{\rho\sigma\mu} = \partial_\sigma g_{\mu\rho} + \partial_\mu g_{\sigma\rho} - \partial_\rho g_{\sigma\mu}$. Notice that the Rotation coefficients
manifestly couple only to the components of the spin which are orthogonal to the motion.

The spin-induced multipole interactions of black holes
(i.e., neglecting tides) are given by nonminimal couplings
in the action. The electric $\mathcal{I}^L$ and magnetic $\mathcal{J}^L$
multipoles are contracted with the corresponding electric $E_L$ and magnetic
$B_L$ curvature tensors in the Lagrangian,
\begin{equation}
  L_\text{SI} = - \sqrt{-u_\rho u^\rho} \sum_{\ell=2}^\infty \frac{1}{\ell!} (\mathcal{I}^L E_L - \mathcal{J}^L B_L) ,
\end{equation}
see \cite{Goldberger:2009qd, Ross:2012fc}, where now $L$ is a 4-dimensional multi-index.
The curvature tensors are defined by\footnote{One can equivalently use the Weyl
  tensor instead of the Riemann tensor here. The difference corresponds to a redefinition
  of the metric in the full action \cite{Goldberger:2004jt}.}
\begin{align}
  E_{\mu_1 \dots \mu_\ell} &=
  \mathcal{P}_{\mu_1}^{\nu_1} \dots \mathcal{P}_{\mu_{\ell-2}}^{\nu_{\ell-2}}
  \nabla_{(\nu_1} \dots \nabla_{\nu_{\ell-2}}
  R_{\mu_{\ell-1}}{}^{\alpha}{}_{\mu_{\ell})}{}^{\beta} U_\alpha U_\beta , \nnm\\
  B_{\mu_1 \dots \mu_\ell} &=
  \mathcal{P}_{\mu_1}^{\nu_1} \dots \mathcal{P}_{\mu_{\ell-2}}^{\nu_{\ell-2}}
  \nabla_{(\nu_1} \dots \nabla_{\nu_{\ell-2}}
  *\!R_{\mu_{\ell-1}}{}^{\alpha}{}_{\mu_{\ell})}{}^{\beta} U_\alpha U_\beta ,
\end{align}
with the dual to the curvature $*R_{\mu\nu\rho\sigma} = \frac{1}{2} \eta_{\mu\nu}{}^{\alpha\beta} R_{\alpha\beta\rho\sigma}$,
the projector $\mathcal{P}_\mu^\nu = \delta_\mu^\nu + U_\mu U^\nu$, the covariant derivative
$\nabla_\mu$, and our convention for the Riemann tensor is
\begin{equation}
R^{\mu}{}_{\nu\alpha\beta} = \Gamma^{\mu}{}_{\nu \beta , \alpha}
	- \Gamma^{\mu}{}_{\nu \alpha , \beta}
	+ \Gamma^{\rho}{}_{\nu \beta} \Gamma^{\mu}{}_{\rho \alpha}
	- \Gamma^{\rho}{}_{\nu \alpha} \Gamma^{\mu}{}_{\rho \beta} .
\end{equation}
Generalizing Eq.~(\ref{IJBH}) by replacing $a^i$ with $a^\mu$,
we then arrive at the Lagrangian
\begin{equation}\label{LSI}
  L_\text{SI} = - m \sqrt{-u_\rho u^\rho} \sum_{\ell=2}^\infty \Re\left[ \frac{i^\ell a^L}{\ell!} (E_L + i B_L) \right] ,
\end{equation}
where $\Re$ denotes the real part. An alternative way to arrive at this Lagrangian is
the following. By constructing all couplings between spin and curvature
consistent with the symmetries of the problem and to linear order in the curvature,
one arrives at an expression analogous to Eq.~(\ref{LSI}) but with undetermined coefficients
on each interaction term \cite{Levi:Steinhoff:2015:1}. Since each coefficient contributes
to a distinct term in the final potential for the binary motion and since this final potential
is correct in the limit of a test-mass moving in the Kerr geometry, see
Sec.~\ref{sec:test}, all of these coefficients were fixed correctly for the case
of a Kerr black hole in Eq.~(\ref{LSI}).

\subsection{Linear approximation}
In this section, we discuss the weak-field approximation to linear order
in the metric perturbation $h_{\mu\nu} = g_{\mu\nu} - \eta_{\mu\nu}$, where $\eta_{\mu\nu}$
is the Minkowski metric.
Indices are pulled using the Minkowski metric from now on.

For simplicity, we choose the tetrad to be the symmetric matrix square-root of the metric 
\begin{equation}\label{tgauge}
e_a{}^{\mu} = \delta_a^\mu - \frac{1}{2} h_a{}^\mu + \Order(h^2) ,
\end{equation}
which fixes the six tetrad gauge degrees of freedom.
It is also useful to fix the gauges of the point-particle action at this point.
For the worldline gauge we choose $\sigma = t$. Then it hold
$u^0 = 1$, $u^i  = d z^i / d t = v^i$, and
\begin{equation}
  \gamma = \frac{1}{\sqrt{1 - \vct{v}^2}} = \frac{1}{\sqrt{-u_\rho u^\rho}} + \Order(h) .
\end{equation}
We denote time derivatives by a dot from now on,
\begin{equation}
  \dot{~} = \frac{d}{dt} = \frac{d}{d \sigma} = u^\mu \partial_\mu .
\end{equation}
For the spin gauge we choose $\Lambda_0{}^b = \delta_0^b$, so that $\Lambda_a{}^0 = \delta_a^0$ and
from Eq.~(\ref{SSC}) also
\begin{equation}
S_{i0} = - \frac{1}{2} S_{ij} v^j + \Order(h) .
\end{equation}
See Refs.~\cite{Levi:Steinhoff:2015:1, Steinhoff:2015} for details.
In these equations and in the following, spatial and temporal indices
on the spin $S^{ij}$ and on $\Lambda^{ij}$ are in the local frame, while other variables
are given in the coordinate frame.

The curvature tensor reads in the linear approximation
\begin{multline}
  2 R_{\mu\nu\alpha\beta} = \partial_\nu \partial_\alpha h_{\mu\beta} + \partial_\mu \partial_\beta h_{\nu\alpha} - \partial_\mu \partial_\alpha h_{\nu\beta} - \partial_\nu \partial_\beta h_{\mu\alpha} \\
  + \Order(h^2) .
\end{multline}
It follows straightforwardly that
\begin{align}
  E_L &= - \frac{1}{2} \partial_L h_{\mu\nu} U^\mu U^\nu + \Order(\dot h, h^2) , \\
B_{L\mu} &= \frac{1}{2} U^\nu \eta_{\nu(\mu}{}^{\alpha\beta} \partial_{L)} \partial_\alpha h_{\beta\rho} U^\rho + \Order(\dot h, h^2) .
\end{align}
Notice that we can drop time derivatives of $h_{\mu\nu}$ (and the tetrad) in the linear
approximation of the particle action, since these can be partially integrated
to give negligible terms containing $\dot u^\mu = \Order(h)$ or $\dot a^\mu = \Order(h)$.
(Insertion of equations of motion into the action is justified here
since it is equivalent to a variable redefinition \cite{Damour:1990jh}.)
These formulas formally only hold for $\ell \geq 2$, but they are also
correct for
\begin{align}
E &= - \frac{1}{2} h_{\mu\nu} U^\mu U^\nu ,
& B_\mu &= *\omega_\nu{}^{ab} U^\nu U_a e_{b\mu} ,
\end{align}
due to our assumption on the tetrad gauge (\ref{tgauge})\footnote{In fact,
this assumption can be relaxed considerably, we just need that time derivatives
of the tetrad can be dropped. This is the case if the tetrad is
expandable in terms of the metric perturbation only (e.g., no explicit coordinate dependence),
which holds for most tetrad gauges.}.
This allows us to extend the summation over $\ell$ in Eq.~(\ref{LSI})
to $\ell=0,1$. We finally arrive at
\begin{multline}\label{SBHh}
    \mathcal{S}_\text{BH} = \int dt \, \bigg\{
    - \frac{m}{\gamma} + \frac{1}{2} S^{ij} \left[ \Lambda^{ki} \dot{\Lambda}^{k j}
      + v^i \dot{v}^j \right] \\
    + \sum_{\ell=0}^\infty \frac{m U^\nu U^\rho}{2 \gamma \, \ell!} \Re\bigg[
    (i a^\sigma \partial_\sigma )^\ell h_{\nu\rho} \\
    + a^\mu \eta_{\nu\mu}{}^{\alpha\beta} \partial_\alpha (i a^\sigma \partial_\sigma )^{\ell-1} h_{\beta\rho}
    \bigg]
    \bigg\} \\
    + \Order(h^2) .
  \end{multline}
Notice that $\dot{v}^j = \Order(h)$.

For completeness, the gravitational action in the linear approximation and in
harmonic gauge $P^{\mu\nu\alpha\beta} \partial_\mu h_{\alpha\beta} = 0$ reads
\begin{equation}
\mathcal{S}_G = - \frac{1}{64 \pi} \int d^4x \, \partial_\rho h_{\mu\nu} P^{\mu\nu\alpha\beta} \partial^\rho h_{\alpha\beta} + \Order(h^3) ,
\end{equation}
where
\begin{equation}
  P^{\mu\nu\alpha\beta} = \frac{1}{2} (\eta^{\mu\alpha} \eta^{\nu\beta} + \eta^{\nu\alpha} \eta^{\mu\beta} - \eta^{\mu\nu} \eta^{\alpha\beta} ) .
\end{equation}
Note that field equations in the linear approximation correspond to a quadratic
approximation in the field part of the action.

\subsection{Post-Newtonian approximation}
The PN approximation is a weak-field and slow-motion
approximation. The PN leading order is obtained by further
specializing the weak-field approximation from the last section
to leading order in the velocity. It is useful to introduce
a decomposition of the metric perturbation in terms of the fields
$\phi$, $A_i$, and $\sigma_{ij} = \sigma_{ji}$ as
\begin{align}
  h_{00} &= - 2 \phi, & h_{0i} &= - A_i & h_{ij} &= - 2 \phi \delta_{ij} + \sigma_{ij} .
\end{align}
Here $\phi$ is the gravito-electric field (Newtonian potential)
and $A_i$ is the gravito-magnetic field.

The PN action is obtained by removing the metric from the full
action $\mathcal{S}_G + \mathcal{S}_\text{BH1} + \mathcal{S}_\text{BH2}$ for the black-hole binary. For this purpose,
one can obtain the field equations for the metric components by
varying the action, solve the field equations, and insert this
solution into the full action, see e.g.~\cite{Damour:1995kt, Bernard:2015njp}.
This is referred to as the Fokker-action approach.
A different method is to integrate out the field using standard
quantum field theory techniques. That is, obtain the Feynman rules
from the action and evaluate all Feynman diagrams which are nonzero in
the classical limit, see e.g.~\cite{Goldberger:2004jt, Rothstein:2014sra}.
We present the latter approach in detail here, but both are not
difficult at the considered order.

\begin{figure}
\subfloat[][]{\label{diael}
\includegraphics{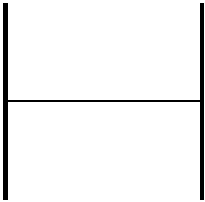}}
\hspace{1.5cm}
\subfloat[][]{\label{diamag}
\includegraphics{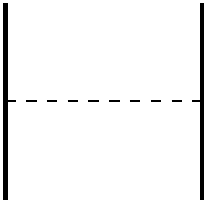}}
\caption{Feynman diagrams of the leading-order gravito-electric and gravito-magnetic interactions.\label{diagrams}}
\end{figure}

The Feynman rule for the interaction of the black-hole
worldline (fat solid line) with a gravito-electric
field $\phi$ (thin solid line) follows directly from
expanding Eq.~(\ref{SBHh}) in $\phi$ leading to
\begin{align}
\parbox{5mm}{\includegraphics{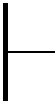}}
  &\approx - m \int dt \left[ \cos(\vct{a} \cdot \vct{\nabla})
    + 2 \frac{\sin(\vct{a} \cdot \vct{\nabla})}{\vct{a} \cdot \vct{\nabla}} (\vct{a} \times \vct{\nabla}) \cdot \vct{v} \right] \phi , \nonumber \\
  &= - m \int dt \left[ \cosh(\vct{a} \times \vct{\nabla})
    + 2 \sinh(\vct{a} \times \vct{\nabla}) \cdot \vct{v} \right] \phi , \label{Feynphi}
\end{align}
in the linear approximation and to linear order in $\vct{v}$.
Here $\vct{\nabla}$ is the partial 3-dimensional derivative.  Note that the hyperbolic sine of a vector is a vector, while the
hyperbolic cosine of a vector is a scalar---it is understood that pairs of adjacent factors of $\bs a\times\bs\nabla$ are contracted, with the grouping of the pairs being inconsequential, as in \mbox{$(\bs a\times\bs\nabla)^3=\bs a\times\bs\nabla(\bs a\times\bs\nabla)^2=(\bs a\times\bs\nabla)^2\bs a\times\bs\nabla$}.  We have also used that
$(\vct{a} \times \vct{\nabla})^2 = \vct{a}^2 \vct{\nabla}^2 - (\vct{a} \cdot \vct{\nabla})^2$ and that $\vct{\nabla}^2 \phi$ can be approximately
removed by a field redefinition in the action; it holds
$\vct{\nabla}^2 \phi = 0$ at the location of the black hole if $\phi$
is sourced by the other back hole.
The Feynman rule (\ref{Feynphi}) directly encodes the source of the field
equation for $\phi$.
The interaction with the gravito-magnetic
field $A_i$ (dashed line) in Eq.~(\ref{SBHh}) is given by
\begin{align}
\parbox{5mm}{\includegraphics{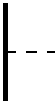}}
  &\approx - m \int dt \left[ \frac{\sin(\vct{a} \cdot \vct{\nabla})}{2 \vct{a} \cdot \vct{\nabla}} \vct{a} \times \vct{\nabla}
    + \cos(\vct{a} \cdot \vct{\nabla}) \vct{v} \right] \cdot \vct{A} , \nonumber \\
  &= - m \int dt \left[ \frac{1}{2} \sinh(\vct{a} \times \vct{\nabla})
    + \cosh(\vct{a} \times \vct{\nabla}) \vct{v} \right] \cdot \vct{A} .
\end{align}
Similarly, for $\sigma_{ij}$ one sees that it only contributes to
quadratic order in $\vct{v}$ and can be neglected here.
The leading-order Feynman diagrams are therefore given
by Fig.~\ref{diagrams}.

The gravitational action reads
\begin{equation}
  \mathcal{S}_G \approx \int \frac{d^4x}{32\pi} \left[ 4 \phi \vct{\nabla}^2 \phi - A_i \vct{\nabla}^2 A_i \right] ,
\end{equation}
from which the Feyman rules for the field propagators
follow as
\begin{align}
\parbox{10mm}{\includegraphics{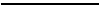}}
  &= \langle \phi(\vct{z}_1, t_1) \phi(\vct{z}_2, t_2) \rangle , \\
  &= \frac{\delta(t_1-t_2)}{R} ,
\end{align}
where $R=|\vct R|$, $\vct R=\vct{z}_1- \vct{z}_2$, and
\begin{align}
\parbox{10mm}{\includegraphics{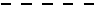}}
  &= \langle A_i(\vct{z}_1, t_1) A_j(\vct{z}_2, t_2) \rangle , \\
  &= -4 \delta_{ij} \frac{\delta(t_1-t_2)}{R} .
\end{align}
The propagators are essentially the (time-symmetric) Green's functions that would be used in
the Fokker-action approach to solve the field equations.

Figure \ref{diael} translates into the following gravito-electric contribution to the interaction potential,
\begin{widetext}
\begin{equation}
  V_\text{el} = - \left[ \frac{1}{2} \cosh(\vct{a}_1 \times \vct{\nabla}\!_1) \cosh(\vct{a}_2 \times \vct{\nabla}\!_2)
    + 2 \vct{v}_1 \cdot \sinh(\vct{a}_1 \times \vct{\nabla}\!_1) \cosh(\vct{a}_2 \times \vct{\nabla}\!_2) \right] \frac{m_1 m_2}{R} + (1 \leftrightarrow 2) ,
\end{equation}
where $\vct{\nabla}\!_A = \partial / \partial \vct{z}_A$ and it was taken into account that the potential
enters the action with a minus sign.
This potential can be obtained in the Fokker-action approach by
collecting all contributions coming from inserting a solution for $\phi$ into
the full action. Similarly, for the gravito-magnetic interaction
(Fig.\ \ref{diamag}) one obtains
\begin{equation}
  V_\text{mag} = \left[ \frac{1}{2} \sinh(\vct{a}_1 \times \vct{\nabla}\!_1)\cdot \sinh(\vct{a}_2 \times \vct{\nabla}\!_2)
    + 2 \vct{v}_1 \cdot \sinh(\vct{a}_2 \times \vct{\nabla}\!_2) \cosh(\vct{a}_1 \times \vct{\nabla}\!_1) \right] \frac{m_1 m_2}{R} + (1 \leftrightarrow 2) .
\end{equation}
The sum of all contributions to the potential reads
\begin{equation}
  V = V_\text{kin} + V_\text{el} + V_\text{mag} = V_\text{kin} - \left[ \cosh(\vct{a}_0 \times \vct{\nabla}) + 2 (\vct{v}_1 - \vct{v}_2) \cdot \sinh(\vct{a}_0 \times \vct{\nabla}) \right] \frac{m_1 m_2}{R} ,
\end{equation}
where we have used $\vct{\nabla}_2 f = - \vct{\nabla}_1 f\equiv -\bs\nabla f$ for any $f(\bs R)$, and the addition theorems for the hyperbolic trigonometric
functions together with the definition (\ref{a0}).  The kinematic
contribution from the last term in the first
line of Eq.~(\ref{SBHh}) is
\begin{align}
  V_\text{kin} &= - \frac{1}{2} S_1^{ij} v_1^i \dot{v}_1^j - \frac{1}{2} S_2^{ij} v_2^i \dot{v}_2^j
                 = \frac{1}{2} \left[ \vct{v}_1 \times \vct{a}_1 - \vct{v}_2 \times \vct{a}_2 \right]
                 \cdot \vct{\nabla} \cosh(\vct{a}_0 \times \vct{\nabla}) \frac{m_1 m_2}{R} ,
\end{align}
\end{widetext}
having inserted the leading-order (even-in-spin) accelerations in the second equality (which induces variable redefinitions \cite{Damour:1990jh}).
The full action finally reads 
\begin{equation}
  \mathcal{S}_\text{LO,PN} = \int dt \left[ \sum_{A=1,2}\left(\frac{m_A}{2}\bs v_A^2+\frac{1}{2} S_A^{ij}  \Lambda_A^{ki} \dot{\Lambda}_A^{k j}\right)
    - V \right],
\end{equation}
where we have dropped the constant rest mass terms.
The Hamiltonian (\ref{Hfinal}) follows from a Legendre transformation,
a specialization to the system's center-of-mass frame (which is achieved simply by using the LO relation $m_1\bs z_1+m_2\bs z_2=0$), and the transition
between hyperbolic and ordinary trigonometric functions as used above.

\subsection{Connection to the test-body limits}\label{sec:test}
In this section we discuss the connections
between the LO PN result and the two test-body limits: test-black-hole motion and geodesic motion in a Kerr spacetime.

Firstly, as pointed out in \cite{Harte:2016vwo}, the \emph{exact} Kerr metric with mass $M$ and spin $M\bs a$ can be written as
\be\label{exactKerr}
g^\mr{Kerr}_{\mu\nu}=\eta_{\mu\nu}+h_{\mu\nu}+2\doe_{(\mu}\xi_{\nu)},
\ee
where $\eta_{\mu\nu}$ is the flat Minkowski metric with connection $\doe_\mu$, the ``gauge vector'' $\xi_\mu$ is given by Eqs.~(65)--(67) of \cite{Harte:2016vwo} (with some notational differences for the spatial coordinates), and the linearized harmonic-gauge ``perturbation'' $h_{\mu\nu}$ is given (exactly) by
\be\label{hKerr}
h_{00} =- 2 \phi, \quad h_{0i} = - A_i,\quad h_{ij} = - 2 \phi \delta_{ij},
\ee
with
\begin{alignat}{3}\label{potsKerr}
\phi&\,=\,-\cosh(\bs a\times\bs\nabla)\frac{M}{R}&&\,=\,-\frac{Mr}{r^2+a^2\cos^2\theta}, 
\nnm\\
\bs A&\,=\,-2\sinh(\bs a\times\bs\nabla)\frac{M}{R}&&\,=\,2\phi\frac{\bs R\times\bs a}{r^2+a^2},
\end{alignat}
where the oblate spheroidal coordinates $(r,\theta,\Phi)$ and Cartesian coordinates $(X,Y,Z)$ on the flat 3-space are related by
\be
X+iY=\sqrt{r^2+a^2}\,\sin\theta\,e^{i\Phi},\quad Z=r\cos\theta,
\ee
[as in (\ref{coords}) with $a_0\to a$].

Because the LO-PN approximation [leading PN order at each order in spin] comprises a subset of the terms in the linearized approximation [$g_{\mu\nu}=\eta_{\mu\nu}+h_{\mu\nu}+\mc O(h^2)$], and because the addition $+2\doe_{(\mu}\xi_{\nu)}$ in (\ref{exactKerr}) takes precisely the form of a linearized gauge transformation, we can simply drop this term for any linear-order or LO-PN calculation, and take $\eta_{\mu\nu}+h_{\mu\nu}$ from (\ref{hKerr})--(\ref{potsKerr}) as our linearized (harmonic-gauge) Kerr metric.

Now, the action for a spinning test black hole in a Kerr metric is given (by construction) by (\ref{SBHh}) with the $h_{\mu\nu}$ there replaced by the stationary Kerr field (\ref{hKerr})--(\ref{potsKerr}) [instead of the dynamical two-body spacetime].  One finds that this leads precisely to the Hamiltonian given by (\ref{Hfinal})--(\ref{potentials}) for a test black hole with mass $\mu$ and spin $\mu\bs\sigma^*$ in a Kerr field with mass $M$ and spin $M\bs\sigma$.  That same Hamiltonian under the mappings (\ref{spinmap}) between $\bs \sigma$, $\bs\sigma^*$ and $\bs a_1$, $\bs a_2$ is the arbitrary-mass-ratio BBH Hamiltonian.  This demonstrates the equivalence (\ref{testmap}) to all orders in spin.

The LO-PN Hamiltonian for geodesics in a Kerr spacetime is found simply by solving $(\eta^{\mu\nu}-h^{\mu\nu})P_\mu P_\nu=-\mu^2$ with $P_\mu=(-H,\bs P)$ for $H$, leading via (\ref{hKerr})--(\ref{potsKerr}) to
\be
H_\mr{geod.}=\mu+\frac{\bs P^2}{2\mu}+\mu\phi+\bs P\cdot\bs A+\mr{NLO}.
\ee
We see that, with the Kerr spin $\bs a\to\bs a_0$, the even-in-spin part matches with that of the BBH Hamiltonian (\ref{Hfinal})--(\ref{potentials}), thus establishing the equivalence (\ref{a0map}) to all orders in spin.  Furthermore, the only addition needed to fix the odd-in-spin part is the last term in (\ref{Hfinal}), \mbox{$\frac{1}{2}\bs P\times\bs\sigma^*\cdot\bs\nabla \phi$}, which is itself expressed very simply in terms of the potential $\phi$ for a rotating black hole with spin $\bs a_0$.  One sees from the derivation in Sec.~\ref{derivation} that this term arises precisely from the kinematic $v^i\dot v^j$ terms in (\ref{SBHh}), from the relativistic top action.

\section{Conclusion}\label{conclude}

We have derived the leading-PN-order, all-orders-in-spin Hamiltonian for a binary black hole, and shown that it bears two distinct intriguing relationships to test-body motion in a Kerr spacetime.

These results are clearly relevant to efforts to develop effective-one-body (EOB) Hamiltonians for BBHs (see e.g.~\cite{Buonanno99,Damour1SEOB,DJS,Barausse:2009xi,Barausse:2011ys,Balmelli:Damour:NLOSS}).  The direct mapping (\ref{testmap}) to the test-body limit (for even and odd parts) is reminiscent of the spinning EOB models of \cite{Barausse:2009xi,Barausse:2011ys} which recover the exact (all-PN-order) test-body limit at linear order in the spin of the test body.  An extension of this strain of EOB models to include higher orders in the test spin would naturally incorporate the new LO-PN arbitrary-spin results, but this would require adding an infinite number of ever more complicated terms to the EOB Hamiltonian---unless one could find a convenient resummation of the arbitrary-test-spin test-black-hole dynamics (which seems less implausible than before in light of the remarkable resummation at LO).  On the other hand, the alternate mapping of (\ref{a0map}) and (\ref{a0}) is a feature of the alternative strain of spinning EOB models including \cite{DJS,Balmelli:Damour:NLOSS}, which are now seen to correctly encode the full even part as in (\ref{a0map}), but not the full odd part.  Thus, both primary strains of EOB models bear some relationship to the new LO arbitrary-spin results, but neither of them as they stand currently encapsulates all of the new results.  

Regardless of any EOB framework, the new ``LO-S$^\infty$'' results reveal nonperturbative mappings of real BBH dynamics to effective test-body motion, and this unexpected fact warrants much further consideration.  It remains to be seen whether these leading-order mappings have analogs at higher post-Newtonian orders---the approach in Ref.~\cite{Guevara:2017csg} could be beneficial to work this out.

\acknowledgements

We thank Alejandro Boh\'e, Alessandra Buonanno, \'Eanna Flanagan, Abraham Harte, Tanja Hinderer, Sylvain Marsat, and two anonymous referees for their comments on this draft and/or for our many fruitful conversations.


\begin{thebibliography}{85}%
\makeatletter
\providecommand \@ifxundefined [1]{%
 \@ifx{#1\undefined}
}%
\providecommand \@ifnum [1]{%
 \ifnum #1\expandafter \@firstoftwo
 \else \expandafter \@secondoftwo
 \fi
}%
\providecommand \@ifx [1]{%
 \ifx #1\expandafter \@firstoftwo
 \else \expandafter \@secondoftwo
 \fi
}%
\providecommand \natexlab [1]{#1}%
\providecommand \enquote  [1]{``#1''}%
\providecommand \bibnamefont  [1]{#1}%
\providecommand \bibfnamefont [1]{#1}%
\providecommand \citenamefont [1]{#1}%
\providecommand \href@noop [0]{\@secondoftwo}%
\providecommand \href [0]{\begingroup \@sanitize@url \@href}%
\providecommand \@href[1]{\@@startlink{#1}\@@href}%
\providecommand \@@href[1]{\endgroup#1\@@endlink}%
\providecommand \@sanitize@url [0]{\catcode `\\12\catcode `\$12\catcode
  `\&12\catcode `\#12\catcode `\^12\catcode `\_12\catcode `\%12\relax}%
\providecommand \@@startlink[1]{}%
\providecommand \@@endlink[0]{}%
\providecommand \url  [0]{\begingroup\@sanitize@url \@url }%
\providecommand \@url [1]{\endgroup\@href {#1}{\urlprefix }}%
\providecommand \urlprefix  [0]{URL }%
\providecommand \Eprint [0]{\href }%
\providecommand \doibase [0]{http://dx.doi.org/}%
\providecommand \selectlanguage [0]{\@gobble}%
\providecommand \bibinfo  [0]{\@secondoftwo}%
\providecommand \bibfield  [0]{\@secondoftwo}%
\providecommand \translation [1]{[#1]}%
\providecommand \BibitemOpen [0]{}%
\providecommand \bibitemStop [0]{}%
\providecommand \bibitemNoStop [0]{.\EOS\space}%
\providecommand \EOS [0]{\spacefactor3000\relax}%
\providecommand \BibitemShut  [1]{\csname bibitem#1\endcsname}%
\let\auto@bib@innerbib\@empty
\bibitem [{\citenamefont {Abbott}\ \emph
  {et~al.}(2016{\natexlab{a}})\citenamefont {Abbott} \emph
  {et~al.}}]{detectionpaper}%
  \BibitemOpen
  \bibfield  {author} {\bibinfo {author} {\bibfnamefont {B.~P.}\ \bibnamefont
  {Abbott}} \emph {et~al.} (\bibinfo {collaboration} {Virgo, LIGO
  Scientific}),\ }\bibfield  {title} {\enquote {\bibinfo {title} {{Observation
  of Gravitational Waves from a Binary Black Hole Merger}},}\ }\href {\doibase
  10.1103/PhysRevLett.116.061102} {\bibfield  {journal} {\bibinfo  {journal}
  {Phys. Rev. Lett.}\ }\textbf {\bibinfo {volume} {116}},\ \bibinfo {pages}
  {061102} (\bibinfo {year} {2016}{\natexlab{a}})},\ \Eprint
  {http://arxiv.org/abs/1602.03837} {arXiv:1602.03837 [gr-qc]} \BibitemShut
  {NoStop}%
\bibitem [{\citenamefont {Abbott}\ \emph
  {et~al.}(2016{\natexlab{b}})\citenamefont {Abbott} \emph
  {et~al.}}]{BoxingDay}%
  \BibitemOpen
  \bibfield  {author} {\bibinfo {author} {\bibfnamefont {B.~P.}\ \bibnamefont
  {Abbott}} \emph {et~al.} (\bibinfo {collaboration} {Virgo, LIGO
  Scientific}),\ }\bibfield  {title} {\enquote {\bibinfo {title} {{GW151226:
  Observation of Gravitational Waves from a 22-Solar-Mass Binary Black Hole
  Coalescence}},}\ }\href {\doibase 10.1103/PhysRevLett.116.241103} {\bibfield
  {journal} {\bibinfo  {journal} {Phys. Rev. Lett.}\ }\textbf {\bibinfo
  {volume} {116}},\ \bibinfo {pages} {241103} (\bibinfo {year}
  {2016}{\natexlab{b}})},\ \Eprint {http://arxiv.org/abs/1606.04855}
  {arXiv:1606.04855 [gr-qc]} \BibitemShut {NoStop}%
\bibitem [{\citenamefont {Abbott}\ \emph {et~al.}(2017)\citenamefont {Abbott}
  \emph {et~al.}}]{Abbott:2017vtc}%
  \BibitemOpen
  \bibfield  {author} {\bibinfo {author} {\bibfnamefont {B.~P.}\ \bibnamefont
  {Abbott}} \emph {et~al.} (\bibinfo {collaboration} {VIRGO, LIGO
  Scientific}),\ }\bibfield  {title} {\enquote {\bibinfo {title} {{GW170104:
  Observation of a 50-Solar-Mass Binary Black Hole Coalescence at Redshift
  0.2}},}\ }\href {\doibase 10.1103/PhysRevLett.118.221101} {\bibfield
  {journal} {\bibinfo  {journal} {Phys. Rev. Lett.}\ }\textbf {\bibinfo
  {volume} {118}},\ \bibinfo {pages} {221101} (\bibinfo {year} {2017})},\
  \Eprint {http://arxiv.org/abs/1706.01812} {arXiv:1706.01812 [gr-qc]}
  \BibitemShut {NoStop}%
\bibitem [{\citenamefont {Abbott}\ \emph
  {et~al.}(2016{\natexlab{c}})\citenamefont {Abbott} \emph
  {et~al.}}]{testingGR}%
  \BibitemOpen
  \bibfield  {author} {\bibinfo {author} {\bibfnamefont {B.~P.}\ \bibnamefont
  {Abbott}} \emph {et~al.} (\bibinfo {collaboration} {Virgo, LIGO
  Scientific}),\ }\bibfield  {title} {\enquote {\bibinfo {title} {{Tests of
  general relativity with GW150914}},}\ }\href {\doibase
  10.1103/PhysRevLett.116.221101} {\bibfield  {journal} {\bibinfo  {journal}
  {Phys. Rev. Lett.}\ }\textbf {\bibinfo {volume} {116}},\ \bibinfo {pages}
  {221101} (\bibinfo {year} {2016}{\natexlab{c}})},\ \Eprint
  {http://arxiv.org/abs/1602.03841} {arXiv:1602.03841 [gr-qc]} \BibitemShut
  {NoStop}%
\bibitem [{\citenamefont {Blanchet}(2014)}]{Blanchet:2006}%
  \BibitemOpen
  \bibfield  {author} {\bibinfo {author} {\bibfnamefont {L.}~\bibnamefont
  {Blanchet}},\ }\bibfield  {title} {\enquote {\bibinfo {title} {{Gravitational
  Radiation from Post-Newtonian Sources and Inspiralling Compact Binaries}},}\
  }\href {\doibase 10.12942/lrr-2014-2} {\bibfield  {journal} {\bibinfo
  {journal} {Living Rev. Rel.}\ }\textbf {\bibinfo {volume} {17}},\ \bibinfo
  {pages} {2} (\bibinfo {year} {2014})},\ \Eprint
  {http://arxiv.org/abs/1310.1528} {arXiv:1310.1528 [gr-qc]} \BibitemShut
  {NoStop}%
\bibitem [{\citenamefont {Poisson}\ and\ \citenamefont
  {Will}(2014)}]{poisson2014gravity}%
  \BibitemOpen
  \bibfield  {author} {\bibinfo {author} {\bibfnamefont {E.}~\bibnamefont
  {Poisson}}\ and\ \bibinfo {author} {\bibfnamefont {C.}~\bibnamefont {Will}},\
  }\href {https://books.google.de/books?id=PZ5cAwAAQBAJ} {\emph {\bibinfo
  {title} {Gravity: Newtonian, Post-Newtonian, Relativistic}}}\ (\bibinfo
  {publisher} {Cambridge University Press},\ \bibinfo {year}
  {2014})\BibitemShut {NoStop}%
\bibitem [{\citenamefont {Detweiler}\ and\ \citenamefont
  {Poisson}(2004)}]{Detweiler:2003ci}%
  \BibitemOpen
  \bibfield  {author} {\bibinfo {author} {\bibfnamefont {S.~L.}\ \bibnamefont
  {Detweiler}}\ and\ \bibinfo {author} {\bibfnamefont {E.}~\bibnamefont
  {Poisson}},\ }\bibfield  {title} {\enquote {\bibinfo {title} {{Low multipole
  contributions to the gravitational selfforce}},}\ }\href {\doibase
  10.1103/PhysRevD.69.084019} {\bibfield  {journal} {\bibinfo  {journal} {Phys.
  Rev.}\ }\textbf {\bibinfo {volume} {D69}},\ \bibinfo {pages} {084019}
  (\bibinfo {year} {2004})},\ \Eprint {http://arxiv.org/abs/gr-qc/0312010}
  {arXiv:gr-qc/0312010 [gr-qc]} \BibitemShut {NoStop}%
\bibitem [{\citenamefont {{Poisson}}\ \emph {et~al.}(2011)\citenamefont
  {{Poisson}}, \citenamefont {{Pound}},\ and\ \citenamefont
  {{Vega}}}]{Poisson:Pound:Vega:2011}%
  \BibitemOpen
  \bibfield  {author} {\bibinfo {author} {\bibfnamefont {E.}~\bibnamefont
  {{Poisson}}}, \bibinfo {author} {\bibfnamefont {A.}~\bibnamefont {{Pound}}},
  \ and\ \bibinfo {author} {\bibfnamefont {I.}~\bibnamefont {{Vega}}},\
  }\bibfield  {title} {\enquote {\bibinfo {title} {{The Motion of Point
  Particles in Curved Spacetime}},}\ }\href {\doibase 10.12942/lrr-2011-7}
  {\bibfield  {journal} {\bibinfo  {journal} {Living Reviews in Relativity}\
  }\textbf {\bibinfo {volume} {14}},\ \bibinfo {pages} {7} (\bibinfo {year}
  {2011})},\ \Eprint {http://arxiv.org/abs/1102.0529} {arXiv:1102.0529 [gr-qc]}
  \BibitemShut {NoStop}%
\bibitem [{\citenamefont {Barack}(2014)}]{Barack:2014}%
  \BibitemOpen
  \bibfield  {author} {\bibinfo {author} {\bibfnamefont {L.}~\bibnamefont
  {Barack}},\ }\bibfield  {title} {\enquote {\bibinfo {title} {Gravitational
  self-force: Orbital mechanics beyond geodesic motion},}\ }in\ \href {\doibase
  10.1007/978-3-319-06349-2_6} {\emph {\bibinfo {booktitle} {General
  Relativity, Cosmology and Astrophysics}}},\ \bibinfo {series} {Fundamental
  Theories of Physics}, Vol.\ \bibinfo {volume} {177}\ (\bibinfo  {publisher}
  {Springer International Publishing},\ \bibinfo {year} {2014})\ pp.\ \bibinfo
  {pages} {147--168}\BibitemShut {NoStop}%
\bibitem [{\citenamefont {Harte}(2015)}]{Harte:review}%
  \BibitemOpen
  \bibfield  {author} {\bibinfo {author} {\bibfnamefont {A.~I.}\ \bibnamefont
  {Harte}},\ }\bibfield  {title} {\enquote {\bibinfo {title} {{Motion in
  classical field theories and the foundations of the self-force problem}},}\
  }\bibfield  {booktitle} {\emph {\bibinfo {booktitle} {{Proceedings, 524th
  WE-Heraeus-Seminar: Equations of Motion in Relativistic Gravity (EOM
  2013)}}},\ }\href {\doibase 10.1007/978-3-319-18335-0_12} {\bibfield
  {journal} {\bibinfo  {journal} {Fund. Theor. Phys.}\ }\textbf {\bibinfo
  {volume} {179}},\ \bibinfo {pages} {327--398} (\bibinfo {year} {2015})},\
  \Eprint {http://arxiv.org/abs/1405.5077} {arXiv:1405.5077 [gr-qc]}
  \BibitemShut {NoStop}%
\bibitem [{\citenamefont {{Pound}}(2015)}]{Pound_review}%
  \BibitemOpen
  \bibfield  {author} {\bibinfo {author} {\bibfnamefont {A.}~\bibnamefont
  {{Pound}}},\ }\bibfield  {title} {\enquote {\bibinfo {title} {{Motion of
  small objects in curved spacetimes: An introduction to gravitational
  self-force}},}\ }\href@noop {} {\bibfield  {journal} {\bibinfo  {journal}
  {ArXiv e-prints}\ } (\bibinfo {year} {2015})},\ \Eprint
  {http://arxiv.org/abs/1506.06245} {arXiv:1506.06245 [gr-qc]} \BibitemShut
  {NoStop}%
\bibitem [{\citenamefont {Bini}\ and\ \citenamefont
  {Damour}(2016)}]{Bini:2016cje}%
  \BibitemOpen
  \bibfield  {author} {\bibinfo {author} {\bibfnamefont {D.}~\bibnamefont
  {Bini}}\ and\ \bibinfo {author} {\bibfnamefont {T.}~\bibnamefont {Damour}},\
  }\bibfield  {title} {\enquote {\bibinfo {title} {{Conservative second-order
  gravitational self-force on circular orbits and the effective one-body
  formalism}},}\ }\href {\doibase 10.1103/PhysRevD.93.104040} {\bibfield
  {journal} {\bibinfo  {journal} {Phys. Rev.}\ }\textbf {\bibinfo {volume}
  {D93}},\ \bibinfo {pages} {104040} (\bibinfo {year} {2016})},\ \Eprint
  {http://arxiv.org/abs/1603.09175} {arXiv:1603.09175 [gr-qc]} \BibitemShut
  {NoStop}%
\bibitem [{\citenamefont {Bini}\ \emph {et~al.}(2016)\citenamefont {Bini},
  \citenamefont {Damour},\ and\ \citenamefont {Geralico}}]{Bini:2016dvs}%
  \BibitemOpen
  \bibfield  {author} {\bibinfo {author} {\bibfnamefont {D.}~\bibnamefont
  {Bini}}, \bibinfo {author} {\bibfnamefont {T.}~\bibnamefont {Damour}}, \ and\
  \bibinfo {author} {\bibfnamefont {A.}~\bibnamefont {Geralico}},\ }\bibfield
  {title} {\enquote {\bibinfo {title} {{High post-Newtonian order gravitational
  self-force analytical results for eccentric equatorial orbits around a Kerr
  black hole}},}\ }\href {\doibase 10.1103/PhysRevD.93.124058} {\bibfield
  {journal} {\bibinfo  {journal} {Phys. Rev.}\ }\textbf {\bibinfo {volume}
  {D93}},\ \bibinfo {pages} {124058} (\bibinfo {year} {2016})},\ \Eprint
  {http://arxiv.org/abs/1602.08282} {arXiv:1602.08282 [gr-qc]} \BibitemShut
  {NoStop}%
\bibitem [{\citenamefont {Damour}\ \emph {et~al.}(1991)\citenamefont {Damour},
  \citenamefont {Soffel},\ and\ \citenamefont {Xu}}]{DSX1}%
  \BibitemOpen
  \bibfield  {author} {\bibinfo {author} {\bibfnamefont {T.}~\bibnamefont
  {Damour}}, \bibinfo {author} {\bibfnamefont {M.}~\bibnamefont {Soffel}}, \
  and\ \bibinfo {author} {\bibfnamefont {C.}~\bibnamefont {Xu}},\ }\bibfield
  {title} {\enquote {\bibinfo {title} {{General-relativistic celestial
  mechanics. I. Method and definition of reference systems}},}\ }\href
  {\doibase 10.1103/PhysRevD.43.3273} {\bibfield  {journal} {\bibinfo
  {journal} {Phys. Rev. D}\ }\textbf {\bibinfo {volume} {43}},\ \bibinfo
  {pages} {3273--3307} (\bibinfo {year} {1991})}\BibitemShut {NoStop}%
\bibitem [{\citenamefont {Damour}\ \emph {et~al.}(1992)\citenamefont {Damour},
  \citenamefont {Soffel},\ and\ \citenamefont {Xu}}]{DSX2}%
  \BibitemOpen
  \bibfield  {author} {\bibinfo {author} {\bibfnamefont {T.}~\bibnamefont
  {Damour}}, \bibinfo {author} {\bibfnamefont {M.}~\bibnamefont {Soffel}}, \
  and\ \bibinfo {author} {\bibfnamefont {C.}~\bibnamefont {Xu}},\ }\bibfield
  {title} {\enquote {\bibinfo {title} {{General-relativistic celestial
  mechanics II. Translational equations of motion}},}\ }\href {\doibase
  10.1103/PhysRevD.45.1017} {\bibfield  {journal} {\bibinfo  {journal} {Phys.
  Rev. D}\ }\textbf {\bibinfo {volume} {45}},\ \bibinfo {pages} {1017--1044}
  (\bibinfo {year} {1992})}\BibitemShut {NoStop}%
\bibitem [{\citenamefont {Racine}\ and\ \citenamefont {Flanagan}(2005)}]{RF}%
  \BibitemOpen
  \bibfield  {author} {\bibinfo {author} {\bibfnamefont {{\'E}.}~\bibnamefont
  {Racine}}\ and\ \bibinfo {author} {\bibfnamefont {{\'E}.~{\'E}.}\
  \bibnamefont {Flanagan}},\ }\bibfield  {title} {\enquote {\bibinfo {title}
  {{Post-1-Newtonian equations of motion for systems of arbitrarily structured
  bodies}},}\ }\href {\doibase 10.1103/PhysRevD.71.044010,
  10.1103/PhysRevD.88.089903} {\bibfield  {journal} {\bibinfo  {journal} {Phys.
  Rev.}\ }\textbf {\bibinfo {volume} {D71}},\ \bibinfo {pages} {044010}
  (\bibinfo {year} {2005})},\ \bibinfo {note} {[Erratum: Phys.
  Rev.D88,no.8,089903(2013)]},\ \Eprint {http://arxiv.org/abs/gr-qc/0404101}
  {arXiv:gr-qc/0404101 [gr-qc]} \BibitemShut {NoStop}%
\bibitem [{\citenamefont {Flanagan}\ and\ \citenamefont
  {Hinderer}(2008)}]{Flanagan:2007ix}%
  \BibitemOpen
  \bibfield  {author} {\bibinfo {author} {\bibfnamefont {{\'E}.~{\'E}.}\
  \bibnamefont {Flanagan}}\ and\ \bibinfo {author} {\bibfnamefont
  {T.}~\bibnamefont {Hinderer}},\ }\bibfield  {title} {\enquote {\bibinfo
  {title} {{Constraining neutron star tidal Love numbers with gravitational
  wave detectors}},}\ }\href {\doibase 10.1103/PhysRevD.77.021502} {\bibfield
  {journal} {\bibinfo  {journal} {Phys. Rev.}\ }\textbf {\bibinfo {volume}
  {D77}},\ \bibinfo {pages} {021502} (\bibinfo {year} {2008})},\ \Eprint
  {http://arxiv.org/abs/0709.1915} {arXiv:0709.1915 [astro-ph]} \BibitemShut
  {NoStop}%
\bibitem [{\citenamefont {Chakrabarti}\ \emph {et~al.}(2013)\citenamefont
  {Chakrabarti}, \citenamefont {Delsate},\ and\ \citenamefont
  {Steinhoff}}]{Chakrabarti:2013xza}%
  \BibitemOpen
  \bibfield  {author} {\bibinfo {author} {\bibfnamefont {S.}~\bibnamefont
  {Chakrabarti}}, \bibinfo {author} {\bibfnamefont {T.}~\bibnamefont
  {Delsate}}, \ and\ \bibinfo {author} {\bibfnamefont {J.}~\bibnamefont
  {Steinhoff}},\ }\bibfield  {title} {\enquote {\bibinfo {title} {{Effective
  action and linear response of compact objects in Newtonian gravity}},}\
  }\href {\doibase 10.1103/PhysRevD.88.084038} {\bibfield  {journal} {\bibinfo
  {journal} {Phys. Rev.}\ }\textbf {\bibinfo {volume} {D88}},\ \bibinfo {pages}
  {084038} (\bibinfo {year} {2013})},\ \Eprint {http://arxiv.org/abs/1306.5820}
  {arXiv:1306.5820 [gr-qc]} \BibitemShut {NoStop}%
\bibitem [{\citenamefont {Goldberger}\ and\ \citenamefont
  {Rothstein}(2006{\natexlab{a}})}]{Goldberger:2005cd}%
  \BibitemOpen
  \bibfield  {author} {\bibinfo {author} {\bibfnamefont {W.~D.}\ \bibnamefont
  {Goldberger}}\ and\ \bibinfo {author} {\bibfnamefont {I.~Z.}\ \bibnamefont
  {Rothstein}},\ }\bibfield  {title} {\enquote {\bibinfo {title} {{Dissipative
  effects in the worldline approach to black hole dynamics}},}\ }\href
  {\doibase 10.1103/PhysRevD.73.104030} {\bibfield  {journal} {\bibinfo
  {journal} {Phys. Rev.}\ }\textbf {\bibinfo {volume} {D73}},\ \bibinfo {pages}
  {104030} (\bibinfo {year} {2006}{\natexlab{a}})},\ \Eprint
  {http://arxiv.org/abs/hep-th/0511133} {arXiv:hep-th/0511133 [hep-th]}
  \BibitemShut {NoStop}%
\bibitem [{\citenamefont {Hinderer}\ \emph {et~al.}(2016)\citenamefont
  {Hinderer} \emph {et~al.}}]{Hinderer:2016eia}%
  \BibitemOpen
  \bibfield  {author} {\bibinfo {author} {\bibfnamefont {T.}~\bibnamefont
  {Hinderer}} \emph {et~al.},\ }\bibfield  {title} {\enquote {\bibinfo {title}
  {{Effects of neutron-star dynamic tides on gravitational waveforms within the
  effective-one-body approach}},}\ }\href {\doibase
  10.1103/PhysRevLett.116.181101} {\bibfield  {journal} {\bibinfo  {journal}
  {Phys. Rev. Lett.}\ }\textbf {\bibinfo {volume} {116}},\ \bibinfo {pages}
  {181101} (\bibinfo {year} {2016})},\ \Eprint
  {http://arxiv.org/abs/1602.00599} {arXiv:1602.00599 [gr-qc]} \BibitemShut
  {NoStop}%
\bibitem [{\citenamefont {Poisson}(1998)}]{Poisson:1997ha}%
  \BibitemOpen
  \bibfield  {author} {\bibinfo {author} {\bibfnamefont {E.}~\bibnamefont
  {Poisson}},\ }\bibfield  {title} {\enquote {\bibinfo {title} {{Gravitational
  waves from inspiraling compact binaries: The Quadrupole moment term}},}\
  }\href {\doibase 10.1103/PhysRevD.57.5287} {\bibfield  {journal} {\bibinfo
  {journal} {Phys. Rev.}\ }\textbf {\bibinfo {volume} {D57}},\ \bibinfo {pages}
  {5287--5290} (\bibinfo {year} {1998})},\ \Eprint
  {http://arxiv.org/abs/gr-qc/9709032} {arXiv:gr-qc/9709032 [gr-qc]}
  \BibitemShut {NoStop}%
\bibitem [{\citenamefont {Hansen}(1974)}]{Hansen:1974zz}%
  \BibitemOpen
  \bibfield  {author} {\bibinfo {author} {\bibfnamefont {R.~O.}\ \bibnamefont
  {Hansen}},\ }\bibfield  {title} {\enquote {\bibinfo {title} {{Multipole
  moments of stationary space-times}},}\ }\href {\doibase 10.1063/1.1666501}
  {\bibfield  {journal} {\bibinfo  {journal} {J. Math. Phys.}\ }\textbf
  {\bibinfo {volume} {15}},\ \bibinfo {pages} {46--52} (\bibinfo {year}
  {1974})}\BibitemShut {NoStop}%
\bibitem [{\citenamefont {Porto}\ and\ \citenamefont
  {Rothstein}(2008{\natexlab{a}})}]{Porto:Rothstein:2008:2}%
  \BibitemOpen
  \bibfield  {author} {\bibinfo {author} {\bibfnamefont {R.~A.}\ \bibnamefont
  {Porto}}\ and\ \bibinfo {author} {\bibfnamefont {I.~Z.}\ \bibnamefont
  {Rothstein}},\ }\bibfield  {title} {\enquote {\bibinfo {title} {{Next to
  Leading Order Spin(1)Spin(1) Effects in the Motion of Inspiralling Compact
  Binaries}},}\ }\href {\doibase 10.1103/PhysRevD.81.029905,
  10.1103/PhysRevD.78.044013} {\bibfield  {journal} {\bibinfo  {journal} {Phys.
  Rev.}\ }\textbf {\bibinfo {volume} {D78}},\ \bibinfo {pages} {044013}
  (\bibinfo {year} {2008}{\natexlab{a}})},\ \bibinfo {note} {[Erratum: Phys.
  Rev.D81,029905(2010)]},\ \Eprint {http://arxiv.org/abs/0804.0260}
  {arXiv:0804.0260 [gr-qc]} \BibitemShut {NoStop}%
\bibitem [{\citenamefont {Taylor}\ and\ \citenamefont
  {Poisson}(2008)}]{Taylor:2008xy}%
  \BibitemOpen
  \bibfield  {author} {\bibinfo {author} {\bibfnamefont {S.}~\bibnamefont
  {Taylor}}\ and\ \bibinfo {author} {\bibfnamefont {E.}~\bibnamefont
  {Poisson}},\ }\bibfield  {title} {\enquote {\bibinfo {title} {{Nonrotating
  black hole in a post-Newtonian tidal environment}},}\ }\href {\doibase
  10.1103/PhysRevD.78.084016} {\bibfield  {journal} {\bibinfo  {journal} {Phys.
  Rev.}\ }\textbf {\bibinfo {volume} {D78}},\ \bibinfo {pages} {084016}
  (\bibinfo {year} {2008})},\ \Eprint {http://arxiv.org/abs/0806.3052}
  {arXiv:0806.3052 [gr-qc]} \BibitemShut {NoStop}%
\bibitem [{\citenamefont {Damour}\ and\ \citenamefont
  {Nagar}(2009)}]{Damour:2009vw}%
  \BibitemOpen
  \bibfield  {author} {\bibinfo {author} {\bibfnamefont {T.}~\bibnamefont
  {Damour}}\ and\ \bibinfo {author} {\bibfnamefont {A.}~\bibnamefont {Nagar}},\
  }\bibfield  {title} {\enquote {\bibinfo {title} {{Relativistic tidal
  properties of neutron stars}},}\ }\href {\doibase 10.1103/PhysRevD.80.084035}
  {\bibfield  {journal} {\bibinfo  {journal} {Phys. Rev.}\ }\textbf {\bibinfo
  {volume} {D80}},\ \bibinfo {pages} {084035} (\bibinfo {year} {2009})},\
  \Eprint {http://arxiv.org/abs/0906.0096} {arXiv:0906.0096 [gr-qc]}
  \BibitemShut {NoStop}%
\bibitem [{\citenamefont {Kol}\ and\ \citenamefont
  {Smolkin}(2012)}]{Kol:2011vg}%
  \BibitemOpen
  \bibfield  {author} {\bibinfo {author} {\bibfnamefont {B.}~\bibnamefont
  {Kol}}\ and\ \bibinfo {author} {\bibfnamefont {M.}~\bibnamefont {Smolkin}},\
  }\bibfield  {title} {\enquote {\bibinfo {title} {{Black hole stereotyping:
  Induced gravito-static polarization}},}\ }\href {\doibase
  10.1007/JHEP02(2012)010} {\bibfield  {journal} {\bibinfo  {journal} {JHEP}\
  }\textbf {\bibinfo {volume} {02}},\ \bibinfo {pages} {010} (\bibinfo {year}
  {2012})},\ \Eprint {http://arxiv.org/abs/1110.3764} {arXiv:1110.3764
  [hep-th]} \BibitemShut {NoStop}%
\bibitem [{\citenamefont {Pani}\ \emph {et~al.}(2015)\citenamefont {Pani},
  \citenamefont {Gualtieri}, \citenamefont {Maselli},\ and\ \citenamefont
  {Ferrari}}]{Pani:2015hfa}%
  \BibitemOpen
  \bibfield  {author} {\bibinfo {author} {\bibfnamefont {P.}~\bibnamefont
  {Pani}}, \bibinfo {author} {\bibfnamefont {L.}~\bibnamefont {Gualtieri}},
  \bibinfo {author} {\bibfnamefont {A.}~\bibnamefont {Maselli}}, \ and\
  \bibinfo {author} {\bibfnamefont {V.}~\bibnamefont {Ferrari}},\ }\bibfield
  {title} {\enquote {\bibinfo {title} {{Tidal deformations of a spinning
  compact object}},}\ }\href {\doibase 10.1103/PhysRevD.92.024010} {\bibfield
  {journal} {\bibinfo  {journal} {Phys. Rev.}\ }\textbf {\bibinfo {volume}
  {D92}},\ \bibinfo {pages} {024010} (\bibinfo {year} {2015})},\ \Eprint
  {http://arxiv.org/abs/1503.07365} {arXiv:1503.07365 [gr-qc]} \BibitemShut
  {NoStop}%
\bibitem [{\citenamefont {Levi}\ and\ \citenamefont
  {Steinhoff}(2016)}]{Levi:Steinhoff:2015:3}%
  \BibitemOpen
  \bibfield  {author} {\bibinfo {author} {\bibfnamefont {M.}~\bibnamefont
  {Levi}}\ and\ \bibinfo {author} {\bibfnamefont {J.}~\bibnamefont
  {Steinhoff}},\ }\bibfield  {title} {\enquote {\bibinfo {title}
  {{Next-to-next-to-leading order gravitational spin-squared potential via the
  effective field theory for spinning objects in the post-Newtonian scheme}},}\
  }\href {\doibase 10.1088/1475-7516/2016/01/008} {\bibfield  {journal}
  {\bibinfo  {journal} {JCAP}\ }\textbf {\bibinfo {volume} {1601}},\ \bibinfo
  {pages} {008} (\bibinfo {year} {2016})},\ \Eprint
  {http://arxiv.org/abs/1506.05794} {arXiv:1506.05794 [gr-qc]} \BibitemShut
  {NoStop}%
\bibitem [{\citenamefont {Damour}\ \emph {et~al.}(2014)\citenamefont {Damour},
  \citenamefont {Jaranowski},\ and\ \citenamefont
  {Sch{\"a}fer}}]{Damour:2014jta}%
  \BibitemOpen
  \bibfield  {author} {\bibinfo {author} {\bibfnamefont {T.}~\bibnamefont
  {Damour}}, \bibinfo {author} {\bibfnamefont {P.}~\bibnamefont {Jaranowski}},
  \ and\ \bibinfo {author} {\bibfnamefont {G.}~\bibnamefont {Sch{\"a}fer}},\
  }\bibfield  {title} {\enquote {\bibinfo {title} {{Nonlocal-in-time action for
  the fourth post-Newtonian conservative dynamics of two-body systems}},}\
  }\href {\doibase 10.1103/PhysRevD.89.064058} {\bibfield  {journal} {\bibinfo
  {journal} {Phys. Rev.}\ }\textbf {\bibinfo {volume} {D89}},\ \bibinfo {pages}
  {064058} (\bibinfo {year} {2014})},\ \Eprint {http://arxiv.org/abs/1401.4548}
  {arXiv:1401.4548 [gr-qc]} \BibitemShut {NoStop}%
\bibitem [{\citenamefont {Damour}\ \emph {et~al.}(2015)\citenamefont {Damour},
  \citenamefont {Jaranowski},\ and\ \citenamefont
  {Sch{\"a}fer}}]{Damour:2015isa}%
  \BibitemOpen
  \bibfield  {author} {\bibinfo {author} {\bibfnamefont {T.}~\bibnamefont
  {Damour}}, \bibinfo {author} {\bibfnamefont {P.}~\bibnamefont {Jaranowski}},
  \ and\ \bibinfo {author} {\bibfnamefont {G.}~\bibnamefont {Sch{\"a}fer}},\
  }\bibfield  {title} {\enquote {\bibinfo {title} {{Fourth post-Newtonian
  effective one-body dynamics}},}\ }\href {\doibase 10.1103/PhysRevD.91.084024}
  {\bibfield  {journal} {\bibinfo  {journal} {Phys. Rev.}\ }\textbf {\bibinfo
  {volume} {D91}},\ \bibinfo {pages} {084024} (\bibinfo {year} {2015})},\
  \Eprint {http://arxiv.org/abs/1502.07245} {arXiv:1502.07245 [gr-qc]}
  \BibitemShut {NoStop}%
\bibitem [{\citenamefont {Bernard}\ \emph {et~al.}(2016)\citenamefont
  {Bernard}, \citenamefont {Blanchet}, \citenamefont {Boh{\'e}}, \citenamefont
  {Faye},\ and\ \citenamefont {Marsat}}]{Bernard:2015njp}%
  \BibitemOpen
  \bibfield  {author} {\bibinfo {author} {\bibfnamefont {L.}~\bibnamefont
  {Bernard}}, \bibinfo {author} {\bibfnamefont {L.}~\bibnamefont {Blanchet}},
  \bibinfo {author} {\bibfnamefont {A.}~\bibnamefont {Boh{\'e}}}, \bibinfo
  {author} {\bibfnamefont {G.}~\bibnamefont {Faye}}, \ and\ \bibinfo {author}
  {\bibfnamefont {S.}~\bibnamefont {Marsat}},\ }\bibfield  {title} {\enquote
  {\bibinfo {title} {{Fokker action of nonspinning compact binaries at the
  fourth post-Newtonian approximation}},}\ }\href {\doibase
  10.1103/PhysRevD.93.084037} {\bibfield  {journal} {\bibinfo  {journal} {Phys.
  Rev.}\ }\textbf {\bibinfo {volume} {D93}},\ \bibinfo {pages} {084037}
  (\bibinfo {year} {2016})},\ \Eprint {http://arxiv.org/abs/1512.02876}
  {arXiv:1512.02876 [gr-qc]} \BibitemShut {NoStop}%
\bibitem [{\citenamefont {Damour}\ \emph {et~al.}(2016)\citenamefont {Damour},
  \citenamefont {Jaranowski},\ and\ \citenamefont
  {Sch{\"a}fer}}]{Damour:2016abl}%
  \BibitemOpen
  \bibfield  {author} {\bibinfo {author} {\bibfnamefont {T.}~\bibnamefont
  {Damour}}, \bibinfo {author} {\bibfnamefont {P.}~\bibnamefont {Jaranowski}},
  \ and\ \bibinfo {author} {\bibfnamefont {G.}~\bibnamefont {Sch{\"a}fer}},\
  }\bibfield  {title} {\enquote {\bibinfo {title} {{Conservative dynamics of
  two-body systems at the fourth post-Newtonian approximation of general
  relativity}},}\ }\href {\doibase 10.1103/PhysRevD.93.084014} {\bibfield
  {journal} {\bibinfo  {journal} {Phys. Rev.}\ }\textbf {\bibinfo {volume}
  {D93}},\ \bibinfo {pages} {084014} (\bibinfo {year} {2016})},\ \Eprint
  {http://arxiv.org/abs/1601.01283} {arXiv:1601.01283 [gr-qc]} \BibitemShut
  {NoStop}%
\bibitem [{\citenamefont {Bernard}\ \emph {et~al.}(2017)\citenamefont
  {Bernard}, \citenamefont {Blanchet}, \citenamefont {Boh{\'e}}, \citenamefont
  {Faye},\ and\ \citenamefont {Marsat}}]{Bernard:2016wrg}%
  \BibitemOpen
  \bibfield  {author} {\bibinfo {author} {\bibfnamefont {L.}~\bibnamefont
  {Bernard}}, \bibinfo {author} {\bibfnamefont {L.}~\bibnamefont {Blanchet}},
  \bibinfo {author} {\bibfnamefont {A.}~\bibnamefont {Boh{\'e}}}, \bibinfo
  {author} {\bibfnamefont {G.}~\bibnamefont {Faye}}, \ and\ \bibinfo {author}
  {\bibfnamefont {S.}~\bibnamefont {Marsat}},\ }\bibfield  {title} {\enquote
  {\bibinfo {title} {{Energy and periastron advance of compact binaries on
  circular orbits at the fourth post-Newtonian order}},}\ }\href {\doibase
  10.1103/PhysRevD.95.044026} {\bibfield  {journal} {\bibinfo  {journal} {Phys.
  Rev.}\ }\textbf {\bibinfo {volume} {D95}},\ \bibinfo {pages} {044026}
  (\bibinfo {year} {2017})},\ \Eprint {http://arxiv.org/abs/1610.07934}
  {arXiv:1610.07934 [gr-qc]} \BibitemShut {NoStop}%
\bibitem [{\citenamefont {Tulczyjew}(1959)}]{Tulczyjew:1959}%
  \BibitemOpen
  \bibfield  {author} {\bibinfo {author} {\bibfnamefont {W.~M.}\ \bibnamefont
  {Tulczyjew}},\ }\bibfield  {title} {\enquote {\bibinfo {title} {{M}otion of
  multipole particles in general relativity theory},}\ }\href@noop {}
  {\bibfield  {journal} {\bibinfo  {journal} {Acta Phys. Pol.}\ }\textbf
  {\bibinfo {volume} {18}},\ \bibinfo {pages} {393--409} (\bibinfo {year}
  {1959})}\BibitemShut {NoStop}%
\bibitem [{\citenamefont {D'Eath}(1975)}]{D'Eath:1975vw}%
  \BibitemOpen
  \bibfield  {author} {\bibinfo {author} {\bibfnamefont {P.~D.}\ \bibnamefont
  {D'Eath}},\ }\bibfield  {title} {\enquote {\bibinfo {title} {{Interaction of
  two black holes in the slow-motion limit}},}\ }\href {\doibase
  10.1103/PhysRevD.12.2183} {\bibfield  {journal} {\bibinfo  {journal} {Phys.
  Rev.}\ }\textbf {\bibinfo {volume} {D12}},\ \bibinfo {pages} {2183--2199}
  (\bibinfo {year} {1975})}\BibitemShut {NoStop}%
\bibitem [{\citenamefont {Barker}\ and\ \citenamefont
  {O'Connell}(1970)}]{PhysRevD.2.1428}%
  \BibitemOpen
  \bibfield  {author} {\bibinfo {author} {\bibfnamefont {B.~M.}\ \bibnamefont
  {Barker}}\ and\ \bibinfo {author} {\bibfnamefont {R.~F.}\ \bibnamefont
  {O'Connell}},\ }\bibfield  {title} {\enquote {\bibinfo {title} {Derivation of
  the equations of motion of a gyroscope from the quantum theory of
  gravitation},}\ }\href {\doibase 10.1103/PhysRevD.2.1428} {\bibfield
  {journal} {\bibinfo  {journal} {Phys. Rev. D}\ }\textbf {\bibinfo {volume}
  {2}},\ \bibinfo {pages} {1428--1435} (\bibinfo {year} {1970})}\BibitemShut
  {NoStop}%
\bibitem [{\citenamefont {Damour}\ and\ \citenamefont
  {Sch{\"a}fer}(1988)}]{Damour:1988mr}%
  \BibitemOpen
  \bibfield  {author} {\bibinfo {author} {\bibfnamefont {T.}~\bibnamefont
  {Damour}}\ and\ \bibinfo {author} {\bibfnamefont {G.}~\bibnamefont
  {Sch{\"a}fer}},\ }\bibfield  {title} {\enquote {\bibinfo {title} {{Higher
  Order Relativistic Periastron Advances and Binary Pulsars}},}\ }\href
  {\doibase 10.1007/BF02828697} {\bibfield  {journal} {\bibinfo  {journal}
  {Nuovo Cim.}\ }\textbf {\bibinfo {volume} {B101}},\ \bibinfo {pages} {127}
  (\bibinfo {year} {1988})}\BibitemShut {NoStop}%
\bibitem [{\citenamefont {Barker}\ and\ \citenamefont
  {O'Connell}(1975)}]{Barker:1975ae}%
  \BibitemOpen
  \bibfield  {author} {\bibinfo {author} {\bibfnamefont {B.~M.}\ \bibnamefont
  {Barker}}\ and\ \bibinfo {author} {\bibfnamefont {R.~F.}\ \bibnamefont
  {O'Connell}},\ }\bibfield  {title} {\enquote {\bibinfo {title}
  {{Gravitational Two-Body Problem with Arbitrary Masses, Spins, and Quadrupole
  Moments}},}\ }\href {\doibase 10.1103/PhysRevD.12.329} {\bibfield  {journal}
  {\bibinfo  {journal} {Phys. Rev.}\ }\textbf {\bibinfo {volume} {D12}},\
  \bibinfo {pages} {329--335} (\bibinfo {year} {1975})}\BibitemShut {NoStop}%
\bibitem [{\citenamefont {Barker}\ and\ \citenamefont
  {O'Connell}(1979)}]{Barker1979}%
  \BibitemOpen
  \bibfield  {author} {\bibinfo {author} {\bibfnamefont {B.~M.}\ \bibnamefont
  {Barker}}\ and\ \bibinfo {author} {\bibfnamefont {R.~F.}\ \bibnamefont
  {O'Connell}},\ }\bibfield  {title} {\enquote {\bibinfo {title} {The
  gravitational interaction: Spin, rotation, and quantum effects-a review},}\
  }\href {\doibase 10.1007/BF00756587} {\bibfield  {journal} {\bibinfo
  {journal} {General Relativity and Gravitation}\ }\textbf {\bibinfo {volume}
  {11}},\ \bibinfo {pages} {149--175} (\bibinfo {year} {1979})}\BibitemShut
  {NoStop}%
\bibitem [{\citenamefont {{Damour}}(2001)}]{Damour1SEOB}%
  \BibitemOpen
  \bibfield  {author} {\bibinfo {author} {\bibfnamefont {T.}~\bibnamefont
  {{Damour}}},\ }\bibfield  {title} {\enquote {\bibinfo {title} {{Coalescence
  of two spinning black holes: An effective one-body approach}},}\ }\href
  {\doibase 10.1103/PhysRevD.64.124013} {\bibfield  {journal} {\bibinfo
  {journal} {\prd}\ }\textbf {\bibinfo {volume} {64}},\ \bibinfo {pages}
  {124013} (\bibinfo {year} {2001})},\ \Eprint
  {http://arxiv.org/abs/gr-qc/0103018} {gr-qc/0103018} \BibitemShut {NoStop}%
\bibitem [{\citenamefont {Racine}(2008)}]{Racine:2008qv}%
  \BibitemOpen
  \bibfield  {author} {\bibinfo {author} {\bibfnamefont {{\'E}.}~\bibnamefont
  {Racine}},\ }\bibfield  {title} {\enquote {\bibinfo {title} {{Analysis of
  spin precession in binary black hole systems including quadrupole-monopole
  interaction}},}\ }\href {\doibase 10.1103/PhysRevD.78.044021} {\bibfield
  {journal} {\bibinfo  {journal} {Phys. Rev.}\ }\textbf {\bibinfo {volume}
  {D78}},\ \bibinfo {pages} {044021} (\bibinfo {year} {2008})},\ \Eprint
  {http://arxiv.org/abs/0803.1820} {arXiv:0803.1820 [gr-qc]} \BibitemShut
  {NoStop}%
\bibitem [{\citenamefont {{Damour}}\ \emph {et~al.}(2008)\citenamefont
  {{Damour}}, \citenamefont {{Jaranowski}},\ and\ \citenamefont
  {{Sch{\"a}fer}}}]{DJS}%
  \BibitemOpen
  \bibfield  {author} {\bibinfo {author} {\bibfnamefont {T.}~\bibnamefont
  {{Damour}}}, \bibinfo {author} {\bibfnamefont {P.}~\bibnamefont
  {{Jaranowski}}}, \ and\ \bibinfo {author} {\bibfnamefont {G.}~\bibnamefont
  {{Sch{\"a}fer}}},\ }\bibfield  {title} {\enquote {\bibinfo {title}
  {{Effective one body approach to the dynamics of two spinning black holes
  with next-to-leading order spin-orbit coupling}},}\ }\href {\doibase
  10.1103/PhysRevD.78.024009} {\bibfield  {journal} {\bibinfo  {journal}
  {\prd}\ }\textbf {\bibinfo {volume} {78}},\ \bibinfo {eid} {024009} (\bibinfo
  {year} {2008})},\ \Eprint {http://arxiv.org/abs/0803.0915} {arXiv:0803.0915
  [gr-qc]} \BibitemShut {NoStop}%
\bibitem [{\citenamefont {Hergt}\ and\ \citenamefont
  {Sch{\"a}fer}(2008{\natexlab{a}})}]{Hergt:2007ha}%
  \BibitemOpen
  \bibfield  {author} {\bibinfo {author} {\bibfnamefont {S.}~\bibnamefont
  {Hergt}}\ and\ \bibinfo {author} {\bibfnamefont {G.}~\bibnamefont
  {Sch{\"a}fer}},\ }\bibfield  {title} {\enquote {\bibinfo {title}
  {{Higher-order-in-spin interaction Hamiltonians for binary black holes from
  source terms of Kerr geometry in approximate ADM coordinates}},}\ }\href
  {\doibase 10.1103/PhysRevD.77.104001} {\bibfield  {journal} {\bibinfo
  {journal} {Phys. Rev.}\ }\textbf {\bibinfo {volume} {D77}},\ \bibinfo {pages}
  {104001} (\bibinfo {year} {2008}{\natexlab{a}})},\ \Eprint
  {http://arxiv.org/abs/0712.1515} {arXiv:0712.1515 [gr-qc]} \BibitemShut
  {NoStop}%
\bibitem [{\citenamefont {Hergt}\ and\ \citenamefont
  {Sch{\"a}fer}(2008{\natexlab{b}})}]{Hergt:2008jn}%
  \BibitemOpen
  \bibfield  {author} {\bibinfo {author} {\bibfnamefont {S.}~\bibnamefont
  {Hergt}}\ and\ \bibinfo {author} {\bibfnamefont {G.}~\bibnamefont
  {Sch{\"a}fer}},\ }\bibfield  {title} {\enquote {\bibinfo {title}
  {{Higher-order-in-spin interaction Hamiltonians for binary black holes from
  Poincare invariance}},}\ }\href {\doibase 10.1103/PhysRevD.78.124004}
  {\bibfield  {journal} {\bibinfo  {journal} {Phys. Rev.}\ }\textbf {\bibinfo
  {volume} {D78}},\ \bibinfo {pages} {124004} (\bibinfo {year}
  {2008}{\natexlab{b}})},\ \Eprint {http://arxiv.org/abs/0809.2208}
  {arXiv:0809.2208 [gr-qc]} \BibitemShut {NoStop}%
\bibitem [{\citenamefont {Levi}\ and\ \citenamefont
  {Steinhoff}(2015{\natexlab{a}})}]{Levi:Steinhoff:2014:2}%
  \BibitemOpen
  \bibfield  {author} {\bibinfo {author} {\bibfnamefont {M.}~\bibnamefont
  {Levi}}\ and\ \bibinfo {author} {\bibfnamefont {J.}~\bibnamefont
  {Steinhoff}},\ }\bibfield  {title} {\enquote {\bibinfo {title} {{Leading
  order finite size effects with spins for inspiralling compact binaries}},}\
  }\href {\doibase 10.1007/JHEP06(2015)059} {\bibfield  {journal} {\bibinfo
  {journal} {JHEP}\ }\textbf {\bibinfo {volume} {06}},\ \bibinfo {pages} {059}
  (\bibinfo {year} {2015}{\natexlab{a}})},\ \Eprint
  {http://arxiv.org/abs/1410.2601} {arXiv:1410.2601 [gr-qc]} \BibitemShut
  {NoStop}%
\bibitem [{\citenamefont {Vaidya}(2015)}]{Vaidya:2014kza}%
  \BibitemOpen
  \bibfield  {author} {\bibinfo {author} {\bibfnamefont {V.}~\bibnamefont
  {Vaidya}},\ }\bibfield  {title} {\enquote {\bibinfo {title} {{Gravitational
  spin Hamiltonians from the S matrix}},}\ }\href {\doibase
  10.1103/PhysRevD.91.024017} {\bibfield  {journal} {\bibinfo  {journal} {Phys.
  Rev.}\ }\textbf {\bibinfo {volume} {D91}},\ \bibinfo {pages} {024017}
  (\bibinfo {year} {2015})},\ \Eprint {http://arxiv.org/abs/1410.5348}
  {arXiv:1410.5348 [hep-th]} \BibitemShut {NoStop}%
\bibitem [{\citenamefont {Marsat}(2015)}]{Marsat:2014xea}%
  \BibitemOpen
  \bibfield  {author} {\bibinfo {author} {\bibfnamefont {S.}~\bibnamefont
  {Marsat}},\ }\bibfield  {title} {\enquote {\bibinfo {title} {{Cubic order
  spin effects in the dynamics and gravitational wave energy flux of compact
  object binaries}},}\ }\href {\doibase 10.1088/0264-9381/32/8/085008}
  {\bibfield  {journal} {\bibinfo  {journal} {Class. Quant. Grav.}\ }\textbf
  {\bibinfo {volume} {32}},\ \bibinfo {pages} {085008} (\bibinfo {year}
  {2015})},\ \Eprint {http://arxiv.org/abs/1411.4118} {arXiv:1411.4118 [gr-qc]}
  \BibitemShut {NoStop}%
\bibitem [{\citenamefont {Damour}\ and\ \citenamefont
  {Taylor}(1992)}]{Damour:1991rd}%
  \BibitemOpen
  \bibfield  {author} {\bibinfo {author} {\bibfnamefont {T.}~\bibnamefont
  {Damour}}\ and\ \bibinfo {author} {\bibfnamefont {J.~H.}\ \bibnamefont
  {Taylor}},\ }\bibfield  {title} {\enquote {\bibinfo {title} {{Strong field
  tests of relativistic gravity and binary pulsars}},}\ }\href {\doibase
  10.1103/PhysRevD.45.1840} {\bibfield  {journal} {\bibinfo  {journal} {Phys.
  Rev.}\ }\textbf {\bibinfo {volume} {D45}},\ \bibinfo {pages} {1840--1868}
  (\bibinfo {year} {1992})}\BibitemShut {NoStop}%
\bibitem [{\citenamefont {Levi}\ and\ \citenamefont
  {Steinhoff}(2015{\natexlab{b}})}]{Levi:Steinhoff:2015:1}%
  \BibitemOpen
  \bibfield  {author} {\bibinfo {author} {\bibfnamefont {M.}~\bibnamefont
  {Levi}}\ and\ \bibinfo {author} {\bibfnamefont {J.}~\bibnamefont
  {Steinhoff}},\ }\bibfield  {title} {\enquote {\bibinfo {title} {{Spinning
  gravitating objects in the effective field theory in the post-Newtonian
  scheme}},}\ }\href {\doibase 10.1007/JHEP09(2015)219} {\bibfield  {journal}
  {\bibinfo  {journal} {JHEP}\ }\textbf {\bibinfo {volume} {09}},\ \bibinfo
  {pages} {219} (\bibinfo {year} {2015}{\natexlab{b}})},\ \Eprint
  {http://arxiv.org/abs/1501.04956} {arXiv:1501.04956 [gr-qc]} \BibitemShut
  {NoStop}%
\bibitem [{\citenamefont {Vines}\ \emph {et~al.}(2016)\citenamefont {Vines},
  \citenamefont {Kunst}, \citenamefont {Steinhoff},\ and\ \citenamefont
  {Hinderer}}]{VKSH}%
  \BibitemOpen
  \bibfield  {author} {\bibinfo {author} {\bibfnamefont {J.}~\bibnamefont
  {Vines}}, \bibinfo {author} {\bibfnamefont {D.}~\bibnamefont {Kunst}},
  \bibinfo {author} {\bibfnamefont {J.}~\bibnamefont {Steinhoff}}, \ and\
  \bibinfo {author} {\bibfnamefont {T.}~\bibnamefont {Hinderer}},\ }\bibfield
  {title} {\enquote {\bibinfo {title} {{Canonical Hamiltonian for an extended
  test body in curved spacetime: To quadratic order in spin}},}\ }\href
  {\doibase 10.1103/PhysRevD.93.103008} {\bibfield  {journal} {\bibinfo
  {journal} {Phys. Rev.}\ }\textbf {\bibinfo {volume} {D93}},\ \bibinfo {pages}
  {103008} (\bibinfo {year} {2016})},\ \Eprint
  {http://arxiv.org/abs/1601.07529} {arXiv:1601.07529 [gr-qc]} \BibitemShut
  {NoStop}%
\bibitem [{\citenamefont {{Balmelli}}\ and\ \citenamefont
  {{Damour}}(2015)}]{Balmelli:Damour:NLOSS}%
  \BibitemOpen
  \bibfield  {author} {\bibinfo {author} {\bibfnamefont {S.}~\bibnamefont
  {{Balmelli}}}\ and\ \bibinfo {author} {\bibfnamefont {T.}~\bibnamefont
  {{Damour}}},\ }\bibfield  {title} {\enquote {\bibinfo {title} {{New
  effective-one-body Hamiltonian with next-to-leading order spin-spin
  coupling}},}\ }\href {\doibase 10.1103/PhysRevD.92.124022} {\bibfield
  {journal} {\bibinfo  {journal} {\prd}\ }\textbf {\bibinfo {volume} {92}},\
  \bibinfo {eid} {124022} (\bibinfo {year} {2015})},\ \Eprint
  {http://arxiv.org/abs/1509.08135} {arXiv:1509.08135 [gr-qc]} \BibitemShut
  {NoStop}%
\bibitem [{\citenamefont {Mathisson}(1937)}]{Mathisson:1937}%
  \BibitemOpen
  \bibfield  {author} {\bibinfo {author} {\bibfnamefont {M.}~\bibnamefont
  {Mathisson}},\ }\bibfield  {title} {\enquote {\bibinfo {title} {Neue mechanik
  materieller systeme},}\ }\href@noop {} {\bibfield  {journal} {\bibinfo
  {journal} {Acta Physica Polonica}\ }\textbf {\bibinfo {volume} {6}},\
  \bibinfo {pages} {163--200} (\bibinfo {year} {1937})}\BibitemShut {NoStop}%
\bibitem [{\citenamefont {Papapetrou}(1951)}]{Papapetrou:1951pa}%
  \BibitemOpen
  \bibfield  {author} {\bibinfo {author} {\bibfnamefont {A.}~\bibnamefont
  {Papapetrou}},\ }\bibfield  {title} {\enquote {\bibinfo {title} {Spinning
  test-particles in general relativity. i},}\ }\href@noop {} {\bibfield
  {journal} {\bibinfo  {journal} {Proc.\ R.\ Soc.\ London\ A}\ }\textbf
  {\bibinfo {volume} {209}},\ \bibinfo {pages} {248} (\bibinfo {year}
  {1951})}\BibitemShut {NoStop}%
\bibitem [{\citenamefont {Dixon}(1979)}]{Dixon:1979}%
  \BibitemOpen
  \bibfield  {author} {\bibinfo {author} {\bibfnamefont {W.~G.}\ \bibnamefont
  {Dixon}},\ }\bibfield  {title} {\enquote {\bibinfo {title} {Extended bodies
  in general relativity: {T}heir description and motion},}\ }in\ \href@noop {}
  {\emph {\bibinfo {booktitle} {Proceedings of the International School of
  Physics Enrico Fermi LXVII}}},\ \bibinfo {editor} {edited by\ \bibinfo
  {editor} {\bibfnamefont {J.}~\bibnamefont {Ehlers}}}\ (\bibinfo  {publisher}
  {North Holland},\ \bibinfo {address} {Amsterdam},\ \bibinfo {year} {1979})\
  pp.\ \bibinfo {pages} {156--219}\BibitemShut {NoStop}%
\bibitem [{\citenamefont {Barausse}\ \emph {et~al.}(2009)\citenamefont
  {Barausse}, \citenamefont {Racine},\ and\ \citenamefont
  {Buonanno}}]{Barausse:Racine:Buonanno:2009}%
  \BibitemOpen
  \bibfield  {author} {\bibinfo {author} {\bibfnamefont {E.}~\bibnamefont
  {Barausse}}, \bibinfo {author} {\bibfnamefont {{\'E}.}~\bibnamefont
  {Racine}}, \ and\ \bibinfo {author} {\bibfnamefont {A.}~\bibnamefont
  {Buonanno}},\ }\bibfield  {title} {\enquote {\bibinfo {title} {{H}amiltonian
  of a spinning test-particle in curved spacetime},}\ }\href {\doibase
  10.1103/PhysRevD.80.104025} {\bibfield  {journal} {\bibinfo  {journal} {Phys.
  Rev. D}\ }\textbf {\bibinfo {volume} {80}},\ \bibinfo {pages} {104025}
  (\bibinfo {year} {2009})},\ \Eprint {http://arxiv.org/abs/0907.4745}
  {arXiv:0907.4745 [gr-qc]} \BibitemShut {NoStop}%
\bibitem [{\citenamefont {Hanson}\ and\ \citenamefont
  {Regge}(1974)}]{Hanson:Regge:1974}%
  \BibitemOpen
  \bibfield  {author} {\bibinfo {author} {\bibfnamefont {A.}~\bibnamefont
  {Hanson}}\ and\ \bibinfo {author} {\bibfnamefont {T.}~\bibnamefont {Regge}},\
  }\bibfield  {title} {\enquote {\bibinfo {title} {The relativistic spherical
  top},}\ }\href {\doibase http://dx.doi.org/10.1016/0003-4916(74)90046-3}
  {\bibfield  {journal} {\bibinfo  {journal} {Annals of Physics}\ }\textbf
  {\bibinfo {volume} {87}},\ \bibinfo {pages} {498 -- 566} (\bibinfo {year}
  {1974})}\BibitemShut {NoStop}%
\bibitem [{\citenamefont {Bailey}\ and\ \citenamefont
  {Israel}(1975)}]{Bailey1975}%
  \BibitemOpen
  \bibfield  {author} {\bibinfo {author} {\bibfnamefont {I.}~\bibnamefont
  {Bailey}}\ and\ \bibinfo {author} {\bibfnamefont {W.}~\bibnamefont
  {Israel}},\ }\bibfield  {title} {\enquote {\bibinfo {title} {Lagrangian
  dynamics of spinning particles and polarized media in general relativity},}\
  }\href {\doibase 10.1007/BF01609434} {\bibfield  {journal} {\bibinfo
  {journal} {Communications in Mathematical Physics}\ }\textbf {\bibinfo
  {volume} {42}},\ \bibinfo {pages} {65--82} (\bibinfo {year}
  {1975})}\BibitemShut {NoStop}%
\bibitem [{\citenamefont {Bailey}\ and\ \citenamefont
  {Israel}(1980)}]{Bailey:Israel:1980}%
  \BibitemOpen
  \bibfield  {author} {\bibinfo {author} {\bibfnamefont {I.}~\bibnamefont
  {Bailey}}\ and\ \bibinfo {author} {\bibfnamefont {W.}~\bibnamefont
  {Israel}},\ }\bibfield  {title} {\enquote {\bibinfo {title} {Relativistic
  dynamics of extended bodies and polarized media: {A}n eccentric approach},}\
  }\href {\doibase 10.1016/0003-4916(80)90231-6} {\bibfield  {journal}
  {\bibinfo  {journal} {Ann. Phys. (N.Y.)}\ }\textbf {\bibinfo {volume}
  {130}},\ \bibinfo {pages} {188--214} (\bibinfo {year} {1980})}\BibitemShut
  {NoStop}%
\bibitem [{\citenamefont {Porto}(2006)}]{Porto:2005ac}%
  \BibitemOpen
  \bibfield  {author} {\bibinfo {author} {\bibfnamefont {R.~A.}\ \bibnamefont
  {Porto}},\ }\bibfield  {title} {\enquote {\bibinfo {title} {{Post-Newtonian
  corrections to the motion of spinning bodies in NRGR}},}\ }\href {\doibase
  10.1103/PhysRevD.73.104031} {\bibfield  {journal} {\bibinfo  {journal} {Phys.
  Rev.}\ }\textbf {\bibinfo {volume} {D73}},\ \bibinfo {pages} {104031}
  (\bibinfo {year} {2006})},\ \Eprint {http://arxiv.org/abs/gr-qc/0511061}
  {arXiv:gr-qc/0511061 [gr-qc]} \BibitemShut {NoStop}%
\bibitem [{\citenamefont {Porto}(2008)}]{Porto:2007qi}%
  \BibitemOpen
  \bibfield  {author} {\bibinfo {author} {\bibfnamefont {R.~A.}\ \bibnamefont
  {Porto}},\ }\bibfield  {title} {\enquote {\bibinfo {title} {{Absorption
  effects due to spin in the worldline approach to black hole dynamics}},}\
  }\href {\doibase 10.1103/PhysRevD.77.064026} {\bibfield  {journal} {\bibinfo
  {journal} {Phys. Rev.}\ }\textbf {\bibinfo {volume} {D77}},\ \bibinfo {pages}
  {064026} (\bibinfo {year} {2008})},\ \Eprint {http://arxiv.org/abs/0710.5150}
  {arXiv:0710.5150 [hep-th]} \BibitemShut {NoStop}%
\bibitem [{\citenamefont {Porto}\ and\ \citenamefont
  {Rothstein}(2008{\natexlab{b}})}]{Porto:2008tb}%
  \BibitemOpen
  \bibfield  {author} {\bibinfo {author} {\bibfnamefont {R.~A.}\ \bibnamefont
  {Porto}}\ and\ \bibinfo {author} {\bibfnamefont {I.~Z.}\ \bibnamefont
  {Rothstein}},\ }\bibfield  {title} {\enquote {\bibinfo {title}
  {{Spin(1)Spin(2) Effects in the Motion of Inspiralling Compact Binaries at
  Third Order in the Post-Newtonian Expansion}},}\ }\href {\doibase
  10.1103/PhysRevD.78.044012, 10.1103/PhysRevD.81.029904} {\bibfield  {journal}
  {\bibinfo  {journal} {Phys. Rev.}\ }\textbf {\bibinfo {volume} {D78}},\
  \bibinfo {pages} {044012} (\bibinfo {year} {2008}{\natexlab{b}})},\ \bibinfo
  {note} {[Erratum: Phys. Rev.D81,029904(2010)]},\ \Eprint
  {http://arxiv.org/abs/0802.0720} {arXiv:0802.0720 [gr-qc]} \BibitemShut
  {NoStop}%
\bibitem [{\citenamefont {Levi}(2010{\natexlab{a}})}]{Levi:2008nh}%
  \BibitemOpen
  \bibfield  {author} {\bibinfo {author} {\bibfnamefont {M.}~\bibnamefont
  {Levi}},\ }\bibfield  {title} {\enquote {\bibinfo {title} {{Next to Leading
  Order gravitational Spin1-Spin2 coupling with Kaluza-Klein reduction}},}\
  }\href {\doibase 10.1103/PhysRevD.82.064029} {\bibfield  {journal} {\bibinfo
  {journal} {Phys. Rev.}\ }\textbf {\bibinfo {volume} {D82}},\ \bibinfo {pages}
  {064029} (\bibinfo {year} {2010}{\natexlab{a}})},\ \Eprint
  {http://arxiv.org/abs/0802.1508} {arXiv:0802.1508 [gr-qc]} \BibitemShut
  {NoStop}%
\bibitem [{\citenamefont {Steinhoff}\ \emph {et~al.}(2008)\citenamefont
  {Steinhoff}, \citenamefont {Sch{\"a}fer},\ and\ \citenamefont
  {Hergt}}]{Steinhoff:2008}%
  \BibitemOpen
  \bibfield  {author} {\bibinfo {author} {\bibfnamefont {J.}~\bibnamefont
  {Steinhoff}}, \bibinfo {author} {\bibfnamefont {G.}~\bibnamefont
  {Sch{\"a}fer}}, \ and\ \bibinfo {author} {\bibfnamefont {S.}~\bibnamefont
  {Hergt}},\ }\bibfield  {title} {\enquote {\bibinfo {title} {{ADM canonical
  formalism for gravitating spinning objects}},}\ }\href {\doibase
  10.1103/PhysRevD.77.104018} {\bibfield  {journal} {\bibinfo  {journal} {Phys.
  Rev.}\ }\textbf {\bibinfo {volume} {D77}},\ \bibinfo {pages} {104018}
  (\bibinfo {year} {2008})},\ \Eprint {http://arxiv.org/abs/0805.3136}
  {arXiv:0805.3136 [gr-qc]} \BibitemShut {NoStop}%
\bibitem [{\citenamefont {Levi}(2010{\natexlab{b}})}]{Levi:2010zu}%
  \BibitemOpen
  \bibfield  {author} {\bibinfo {author} {\bibfnamefont {M.}~\bibnamefont
  {Levi}},\ }\bibfield  {title} {\enquote {\bibinfo {title} {{Next to Leading
  Order gravitational Spin-Orbit coupling in an Effective Field Theory
  approach}},}\ }\href {\doibase 10.1103/PhysRevD.82.104004} {\bibfield
  {journal} {\bibinfo  {journal} {Phys. Rev.}\ }\textbf {\bibinfo {volume}
  {D82}},\ \bibinfo {pages} {104004} (\bibinfo {year} {2010}{\natexlab{b}})},\
  \Eprint {http://arxiv.org/abs/1006.4139} {arXiv:1006.4139 [gr-qc]}
  \BibitemShut {NoStop}%
\bibitem [{\citenamefont {Porto}(2010)}]{Porto:2010tr}%
  \BibitemOpen
  \bibfield  {author} {\bibinfo {author} {\bibfnamefont {R.~A.}\ \bibnamefont
  {Porto}},\ }\bibfield  {title} {\enquote {\bibinfo {title} {{Next to leading
  order spin-orbit effects in the motion of inspiralling compact binaries}},}\
  }\href {\doibase 10.1088/0264-9381/27/20/205001} {\bibfield  {journal}
  {\bibinfo  {journal} {Class. Quant. Grav.}\ }\textbf {\bibinfo {volume}
  {27}},\ \bibinfo {pages} {205001} (\bibinfo {year} {2010})},\ \Eprint
  {http://arxiv.org/abs/1005.5730} {arXiv:1005.5730 [gr-qc]} \BibitemShut
  {NoStop}%
\bibitem [{\citenamefont {Perrodin}(2010)}]{Perrodin:2010dy}%
  \BibitemOpen
  \bibfield  {author} {\bibinfo {author} {\bibfnamefont {D.~L.}\ \bibnamefont
  {Perrodin}},\ }\bibfield  {title} {\enquote {\bibinfo {title} {{Subleading
  Spin-Orbit Correction to the Newtonian Potential in Effective Field Theory
  Formalism}},}\ }in\ \href {\doibase 10.1142/9789814374552_0041} {\emph
  {\bibinfo {booktitle} {{On recent developments in theoretical and
  experimental general relativity, astrophysics and relativistic field
  theories. Proceedings, 12th Marcel Grossmann Meeting on General Relativity,
  Paris, France, July 12-18, 2009. Vol. 1-3}}}}\ (\bibinfo {year} {2010})\ pp.\
  \bibinfo {pages} {725--727},\ \Eprint {http://arxiv.org/abs/1005.0634}
  {arXiv:1005.0634 [gr-qc]} \BibitemShut {NoStop}%
\bibitem [{\citenamefont {Porto}(2016)}]{PORTO20161}%
  \BibitemOpen
  \bibfield  {author} {\bibinfo {author} {\bibfnamefont {R.~A.}\ \bibnamefont
  {Porto}},\ }\bibfield  {title} {\enquote {\bibinfo {title} {The effective
  field theorist’s approach to gravitational dynamics},}\ }\href {\doibase
  https://doi.org/10.1016/j.physrep.2016.04.003} {\bibfield  {journal}
  {\bibinfo  {journal} {Physics Reports}\ }\textbf {\bibinfo {volume} {633}},\
  \bibinfo {pages} {1 -- 104} (\bibinfo {year} {2016})},\ \bibinfo {note} {the
  effective field theorist’s approach to gravitational dynamics},\ \Eprint
  {http://arxiv.org/abs/1601.04914} {arXiv:1601.04914 [gr-qc]} \BibitemShut
  {NoStop}%
\bibitem [{\citenamefont {Steinhoff}(2011)}]{Steinhoff:2011}%
  \BibitemOpen
  \bibfield  {author} {\bibinfo {author} {\bibfnamefont {J.}~\bibnamefont
  {Steinhoff}},\ }\bibfield  {title} {\enquote {\bibinfo {title} {Canonical
  formulation of spin in general relativity},}\ }\href {\doibase
  10.1002/andp.201000178} {\bibfield  {journal} {\bibinfo  {journal} {Ann.
  Phys. (Berlin)}\ }\textbf {\bibinfo {volume} {523}},\ \bibinfo {pages}
  {296--353} (\bibinfo {year} {2011})},\ \Eprint
  {http://arxiv.org/abs/1106.4203} {arXiv:1106.4203 [gr-qc]} \BibitemShut
  {NoStop}%
\bibitem [{\citenamefont {{Steinhoff}}(2015)}]{Steinhoff:2015}%
  \BibitemOpen
  \bibfield  {author} {\bibinfo {author} {\bibfnamefont {J.}~\bibnamefont
  {{Steinhoff}}},\ }\bibfield  {title} {\enquote {\bibinfo {title} {{Spin gauge
  symmetry in the action principle for classical relativistic particles}},}\
  }\href@noop {} {\bibfield  {journal} {\bibinfo  {journal} {ArXiv e-prints}\ }
  (\bibinfo {year} {2015})},\ \Eprint {http://arxiv.org/abs/1501.04951}
  {arXiv:1501.04951 [gr-qc]} \BibitemShut {NoStop}%
\bibitem [{\citenamefont {Boh{\'e}}\ \emph {et~al.}(2015)\citenamefont
  {Boh{\'e}}, \citenamefont {Faye}, \citenamefont {Marsat},\ and\ \citenamefont
  {Porter}}]{Bohe:2015ana}%
  \BibitemOpen
  \bibfield  {author} {\bibinfo {author} {\bibfnamefont {A.}~\bibnamefont
  {Boh{\'e}}}, \bibinfo {author} {\bibfnamefont {G.}~\bibnamefont {Faye}},
  \bibinfo {author} {\bibfnamefont {S.}~\bibnamefont {Marsat}}, \ and\ \bibinfo
  {author} {\bibfnamefont {E.~K.}\ \bibnamefont {Porter}},\ }\bibfield  {title}
  {\enquote {\bibinfo {title} {{Quadratic-in-spin effects in the orbital
  dynamics and gravitational-wave energy flux of compact binaries at the 3PN
  order}},}\ }\href {\doibase 10.1088/0264-9381/32/19/195010} {\bibfield
  {journal} {\bibinfo  {journal} {Class. Quant. Grav.}\ }\textbf {\bibinfo
  {volume} {32}},\ \bibinfo {pages} {195010} (\bibinfo {year} {2015})},\
  \Eprint {http://arxiv.org/abs/1501.01529} {arXiv:1501.01529 [gr-qc]}
  \BibitemShut {NoStop}%
\bibitem [{\citenamefont {Israel}(1970)}]{Israel:1970}%
  \BibitemOpen
  \bibfield  {author} {\bibinfo {author} {\bibfnamefont {W.}~\bibnamefont
  {Israel}},\ }\bibfield  {title} {\enquote {\bibinfo {title} {Source of the
  {K}err metric},}\ }\href {\doibase 10.1103/PhysRevD.2.641} {\bibfield
  {journal} {\bibinfo  {journal} {Phys. Rev. D}\ }\textbf {\bibinfo {volume}
  {2}},\ \bibinfo {pages} {641--646} (\bibinfo {year} {1970})}\BibitemShut
  {NoStop}%
\bibitem [{\citenamefont {Will}(2009)}]{Will:2008ys}%
  \BibitemOpen
  \bibfield  {author} {\bibinfo {author} {\bibfnamefont {C.~M.}\ \bibnamefont
  {Will}},\ }\bibfield  {title} {\enquote {\bibinfo {title} {{Carter-like
  constants of the motion in Newtonian gravity and electrodynamics}},}\ }\href
  {\doibase 10.1103/PhysRevLett.102.061101} {\bibfield  {journal} {\bibinfo
  {journal} {Phys. Rev. Lett.}\ }\textbf {\bibinfo {volume} {102}},\ \bibinfo
  {pages} {061101} (\bibinfo {year} {2009})},\ \Eprint
  {http://arxiv.org/abs/0812.0110} {arXiv:0812.0110 [gr-qc]} \BibitemShut
  {NoStop}%
\bibitem [{\citenamefont {Misner}\ \emph {et~al.}(1973)\citenamefont {Misner},
  \citenamefont {Thorne},\ and\ \citenamefont
  {Wheeler}}]{Misner:Thorne:Wheeler:1973}%
  \BibitemOpen
  \bibfield  {author} {\bibinfo {author} {\bibfnamefont {C.~W.}\ \bibnamefont
  {Misner}}, \bibinfo {author} {\bibfnamefont {K.~S.}\ \bibnamefont {Thorne}},
  \ and\ \bibinfo {author} {\bibfnamefont {J.~A.}\ \bibnamefont {Wheeler}},\
  }\href@noop {} {\emph {\bibinfo {title} {Gravitation}}},\ \bibinfo {edition}
  {21st}\ ed.\ (\bibinfo  {publisher} {W. H. Freeman and Company},\ \bibinfo
  {address} {41 Madison Avenue, New York},\ \bibinfo {year} {1973})\BibitemShut
  {NoStop}%
\bibitem [{\citenamefont {Kyrian}\ and\ \citenamefont
  {Semerak}(2007)}]{Kyrian:2007zz}%
  \BibitemOpen
  \bibfield  {author} {\bibinfo {author} {\bibfnamefont {K.}~\bibnamefont
  {Kyrian}}\ and\ \bibinfo {author} {\bibfnamefont {O.}~\bibnamefont
  {Semerak}},\ }\bibfield  {title} {\enquote {\bibinfo {title} {{Spinning test
  particles in a Kerr field}},}\ }\href {\doibase
  10.1111/j.1365-2966.2007.12502.x} {\bibfield  {journal} {\bibinfo  {journal}
  {Mon. Not. Roy. Astron. Soc.}\ }\textbf {\bibinfo {volume} {382}},\ \bibinfo
  {pages} {1922} (\bibinfo {year} {2007})}\BibitemShut {NoStop}%
\bibitem [{\citenamefont {Goldberger}\ and\ \citenamefont
  {Ross}(2010)}]{Goldberger:2009qd}%
  \BibitemOpen
  \bibfield  {author} {\bibinfo {author} {\bibfnamefont {W.~D.}\ \bibnamefont
  {Goldberger}}\ and\ \bibinfo {author} {\bibfnamefont {A.}~\bibnamefont
  {Ross}},\ }\bibfield  {title} {\enquote {\bibinfo {title} {{Gravitational
  radiative corrections from effective field theory}},}\ }\href {\doibase
  10.1103/PhysRevD.81.124015} {\bibfield  {journal} {\bibinfo  {journal} {Phys.
  Rev.}\ }\textbf {\bibinfo {volume} {D81}},\ \bibinfo {pages} {124015}
  (\bibinfo {year} {2010})},\ \Eprint {http://arxiv.org/abs/0912.4254}
  {arXiv:0912.4254 [gr-qc]} \BibitemShut {NoStop}%
\bibitem [{\citenamefont {Ross}(2012)}]{Ross:2012fc}%
  \BibitemOpen
  \bibfield  {author} {\bibinfo {author} {\bibfnamefont {A.}~\bibnamefont
  {Ross}},\ }\bibfield  {title} {\enquote {\bibinfo {title} {{Multipole
  expansion at the level of the action}},}\ }\href {\doibase
  10.1103/PhysRevD.85.125033} {\bibfield  {journal} {\bibinfo  {journal} {Phys.
  Rev.}\ }\textbf {\bibinfo {volume} {D85}},\ \bibinfo {pages} {125033}
  (\bibinfo {year} {2012})},\ \Eprint {http://arxiv.org/abs/1202.4750}
  {arXiv:1202.4750 [gr-qc]} \BibitemShut {NoStop}%
\bibitem [{\citenamefont {Goldberger}\ and\ \citenamefont
  {Rothstein}(2006{\natexlab{b}})}]{Goldberger:2004jt}%
  \BibitemOpen
  \bibfield  {author} {\bibinfo {author} {\bibfnamefont {W.~D.}\ \bibnamefont
  {Goldberger}}\ and\ \bibinfo {author} {\bibfnamefont {I.~Z.}\ \bibnamefont
  {Rothstein}},\ }\bibfield  {title} {\enquote {\bibinfo {title} {{An Effective
  field theory of gravity for extended objects}},}\ }\href {\doibase
  10.1103/PhysRevD.73.104029} {\bibfield  {journal} {\bibinfo  {journal} {Phys.
  Rev.}\ }\textbf {\bibinfo {volume} {D73}},\ \bibinfo {pages} {104029}
  (\bibinfo {year} {2006}{\natexlab{b}})},\ \Eprint
  {http://arxiv.org/abs/hep-th/0409156} {arXiv:hep-th/0409156 [hep-th]}
  \BibitemShut {NoStop}%
\bibitem [{\citenamefont {Damour}\ and\ \citenamefont
  {Schäfer}(1991)}]{Damour:1990jh}%
  \BibitemOpen
  \bibfield  {author} {\bibinfo {author} {\bibfnamefont {T.}~\bibnamefont
  {Damour}}\ and\ \bibinfo {author} {\bibfnamefont {G.}~\bibnamefont
  {Schäfer}},\ }\bibfield  {title} {\enquote {\bibinfo {title} {{Redefinition
  of position variables and the reduction of higher order Lagrangians}},}\
  }\href {\doibase 10.1063/1.529135} {\bibfield  {journal} {\bibinfo  {journal}
  {J. Math. Phys.}\ }\textbf {\bibinfo {volume} {32}},\ \bibinfo {pages}
  {127--134} (\bibinfo {year} {1991})}\BibitemShut {NoStop}%
\bibitem [{\citenamefont {Damour}\ and\ \citenamefont
  {Esposito-Farese}(1996)}]{Damour:1995kt}%
  \BibitemOpen
  \bibfield  {author} {\bibinfo {author} {\bibfnamefont {T.}~\bibnamefont
  {Damour}}\ and\ \bibinfo {author} {\bibfnamefont {G.}~\bibnamefont
  {Esposito-Farese}},\ }\bibfield  {title} {\enquote {\bibinfo {title}
  {{Testing gravity to second post-Newtonian order: A Field theory
  approach}},}\ }\href {\doibase 10.1103/PhysRevD.53.5541} {\bibfield
  {journal} {\bibinfo  {journal} {Phys. Rev.}\ }\textbf {\bibinfo {volume}
  {D53}},\ \bibinfo {pages} {5541--5578} (\bibinfo {year} {1996})},\ \Eprint
  {http://arxiv.org/abs/gr-qc/9506063} {arXiv:gr-qc/9506063 [gr-qc]}
  \BibitemShut {NoStop}%
\bibitem [{\citenamefont {Rothstein}(2014)}]{Rothstein:2014sra}%
  \BibitemOpen
  \bibfield  {author} {\bibinfo {author} {\bibfnamefont {I.~Z.}\ \bibnamefont
  {Rothstein}},\ }\bibfield  {title} {\enquote {\bibinfo {title} {{Progress in
  effective field theory approach to the binary inspiral problem}},}\ }\href
  {\doibase 10.1007/s10714-014-1726-y} {\bibfield  {journal} {\bibinfo
  {journal} {Gen. Rel. Grav.}\ }\textbf {\bibinfo {volume} {46}},\ \bibinfo
  {pages} {1726} (\bibinfo {year} {2014})}\BibitemShut {NoStop}%
\bibitem [{\citenamefont {Harte}\ and\ \citenamefont
  {Vines}(2016)}]{Harte:2016vwo}%
  \BibitemOpen
  \bibfield  {author} {\bibinfo {author} {\bibfnamefont {A.~I.}\ \bibnamefont
  {Harte}}\ and\ \bibinfo {author} {\bibfnamefont {J.}~\bibnamefont {Vines}},\
  }\bibfield  {title} {\enquote {\bibinfo {title} {{Generating exact solutions
  to Einstein's equation using linearized approximations}},}\ }\href {\doibase
  10.1103/PhysRevD.94.084009} {\bibfield  {journal} {\bibinfo  {journal} {Phys.
  Rev.}\ }\textbf {\bibinfo {volume} {D94}},\ \bibinfo {pages} {084009}
  (\bibinfo {year} {2016})},\ \Eprint {http://arxiv.org/abs/1608.04359}
  {arXiv:1608.04359 [gr-qc]} \BibitemShut {NoStop}%
\bibitem [{\citenamefont {Buonanno}\ and\ \citenamefont
  {Damour}(1999)}]{Buonanno99}%
  \BibitemOpen
  \bibfield  {author} {\bibinfo {author} {\bibfnamefont {A.}~\bibnamefont
  {Buonanno}}\ and\ \bibinfo {author} {\bibfnamefont {T.}~\bibnamefont
  {Damour}},\ }\bibfield  {title} {\enquote {\bibinfo {title} {{Effective
  one-body approach to general relativistic two-body dynamics}},}\ }\href
  {\doibase 10.1103/PhysRevD.59.084006} {\bibfield  {journal} {\bibinfo
  {journal} {\prd}\ }\textbf {\bibinfo {volume} {59}},\ \bibinfo {pages}
  {084006} (\bibinfo {year} {1999})},\ \Eprint
  {http://arxiv.org/abs/gr-qc/9811091} {gr-qc/9811091} \BibitemShut {NoStop}%
\bibitem [{\citenamefont {Barausse}\ and\ \citenamefont
  {Buonanno}(2010)}]{Barausse:2009xi}%
  \BibitemOpen
  \bibfield  {author} {\bibinfo {author} {\bibfnamefont {E.}~\bibnamefont
  {Barausse}}\ and\ \bibinfo {author} {\bibfnamefont {A.}~\bibnamefont
  {Buonanno}},\ }\bibfield  {title} {\enquote {\bibinfo {title} {{Improved
  effective-one-body Hamiltonian for spinning black-hole binaries}},}\ }\href
  {\doibase 10.1103/PhysRevD.81.084024} {\bibfield  {journal} {\bibinfo
  {journal} {\prd}\ }\textbf {\bibinfo {volume} {81}},\ \bibinfo {pages}
  {084024} (\bibinfo {year} {2010})},\ \Eprint
  {http://arxiv.org/abs/gr-qc/0912.3517} {gr-qc/0912.3517} \BibitemShut
  {NoStop}%
\bibitem [{\citenamefont {Barausse}\ and\ \citenamefont
  {Buonanno}(2011)}]{Barausse:2011ys}%
  \BibitemOpen
  \bibfield  {author} {\bibinfo {author} {\bibfnamefont {E.}~\bibnamefont
  {Barausse}}\ and\ \bibinfo {author} {\bibfnamefont {A.}~\bibnamefont
  {Buonanno}},\ }\bibfield  {title} {\enquote {\bibinfo {title} {{Extending the
  effective-one-body Hamiltonian of black-hole binaries to include
  next-to-next-to-leading spin-orbit couplings}},}\ }\href {\doibase
  10.1103/PhysRevD.84.104027} {\bibfield  {journal} {\bibinfo  {journal}
  {\prd}\ }\textbf {\bibinfo {volume} {84}},\ \bibinfo {pages} {104027}
  (\bibinfo {year} {2011})},\ \Eprint {http://arxiv.org/abs/gr-qc/1107.2904}
  {gr-qc/1107.2904} \BibitemShut {NoStop}%
\bibitem [{\citenamefont {Guevara}(2017)}]{Guevara:2017csg}%
  \BibitemOpen
  \bibfield  {author} {\bibinfo {author} {\bibfnamefont {A.}~\bibnamefont
  {Guevara}},\ }\bibfield  {title} {\enquote {\bibinfo {title} {{Holomorphic
  Classical Limit for Spin Effects in Gravitational and Electromagnetic
  Scattering}},}\ }\href@noop {} {\  (\bibinfo {year} {2017})},\ \Eprint
  {http://arxiv.org/abs/1706.02314} {arXiv:1706.02314 [hep-th]} \BibitemShut
  {NoStop}%
\end{thebibliography}

%

\end{document}